\documentclass[nonacm,sigconf]{acmart}

\usepackage{mathtools}
\usepackage{multirow}
\usepackage{subfigure}
\usepackage[linesnumbered,ruled,vlined]{algorithm2e}

\SetKwInOut{Input}{Input}
\SetKwInOut{Output}{Output}

\DeclareMathOperator*{\argmax}{arg\,max}

\begin{document}

\title{Fair and Representative Subset Selection from Data Streams}
\titlenote{To appear in the Web conference 2021 (WWW 2021)}

\author{Yanhao Wang}
\orcid{0000-0002-7661-3917}
\affiliation{%
  \institution{University of Helsinki}
  \city{Helsinki}
  \country{Finland}}
\email{yanhao.wang@helsinki.fi}

\author{Francesco Fabbri}
\authornote{This research was done while Francesco Fabbri worked at the University of Helsinki.}
\affiliation{%
  \institution{Pompeu Fabra University \& Eurecat}
  \city{Barcelona}
  \country{Spain}}
\email{francesco.fabbri@upf.edu}

\author{Michael Mathioudakis}
\orcid{0000-0003-0074-3966}
\affiliation{%
  \institution{University of Helsinki}
  \city{Helsinki}
  \country{Finland}}
\email{michael.mathioudakis@helsinki.fi}

\begin{abstract}
  We study the problem of extracting a small subset of representative items from a large data stream.
  In many data mining and machine learning applications such as social network analysis and recommender systems,
  this problem can be formulated as maximizing a monotone submodular function subject to a cardinality constraint $k$.
  In this work, we consider the setting where data items in the stream belong to one of several disjoint groups
  and investigate the optimization problem with an additional \emph{fairness} constraint that limits selection
  to a given number of items from each group. We then propose efficient algorithms for the fairness-aware variant
  of the streaming submodular maximization problem. 
  In particular, we first give a $ (\frac{1}{2}-\varepsilon) $-approximation algorithm that
  requires $ O(\frac{1}{\varepsilon} \log \frac{k}{\varepsilon}) $ passes over the stream
  for any constant $ \varepsilon>0 $. Moreover, we give a single-pass streaming algorithm that
  has the same approximation ratio of $(\frac{1}{2}-\varepsilon)$ when unlimited buffer sizes and
  post-processing time are permitted, and discuss how to adapt it to more practical settings
  where the buffer sizes are bounded.
  Finally, we demonstrate the efficiency and effectiveness of our proposed algorithms on two real-world applications,
  namely \emph{maximum coverage on large graphs} and \emph{personalized recommendation}.
\end{abstract}

\begin{CCSXML}
  <ccs2012>
    <concept>
      <concept_id>10002951.10003227.10003351.10003446</concept_id>
      <concept_desc>Information systems~Data stream mining</concept_desc>
      <concept_significance>500</concept_significance>
    </concept>
  </ccs2012>
\end{CCSXML}

\ccsdesc[500]{Information systems~Data stream mining}

\keywords{algorithmic fairness, approximation algorithm, data summarization, streaming algorithm, submodular maximization}

\maketitle

%!TEX root = main.tex
\section{Introduction}\label{sec:intro}

A crucial task in modern data-driven applications, ranging from influence
maximization~\cite{DBLP:conf/kdd/KempeKT03,DBLP:conf/www/StoicaHC20}
and recommender systems~\cite{DBLP:conf/www/SerbosQMPT17,DBLP:conf/icml/Norouzi-FardTMZ18},
to nonparametric learning~\cite{DBLP:conf/icml/GomesK10,DBLP:conf/kdd/BadanidiyuruMKK14}
and data summarization~\cite{DBLP:conf/sdm/SahaG09,DBLP:conf/www/EpastoLVZ17},
is to extract a few representatives from a large dataset.
In all aforementioned applications, this task can be formulated as selecting a subset of items
to maximize a utility function that quantifies the ``representativeness'' (or ``utility'')
of the selected subset. Oftentimes, the objective function satisfies \emph{submodularity},
a property of ``diminishing returns'' such that adding an item to a smaller set always leads
to a greater increase in utility than adding it to a larger set. Consequently, maximizing
submodular set functions subject to a cardinality constraint (i.e., the size of the selected
subset is limited to an integer $k$) is general enough to model many practical problems
in data mining and machine learning. In this work, we adopt the same formulation for representative
item selection.

The classic approach to the cardinality-constrained submodular maximization problem is the
\textsc{Greedy} algorithm proposed by Nemhauser et al.~\cite{DBLP:journals/mp/NemhauserWF78},
which achieves an approximation factor of $(1-\frac{1}{e})$ that is NP-hard to
improve~\cite{DBLP:journals/jacm/Feige98}. In many real-world scenarios, however, the data become
too large to fit in memory or arrive incrementally at a high rate. In such cases, the \textsc{Greedy}
algorithm becomes very inefficient because it requires $k$ repeated sequential scans over the whole dataset.
Therefore, streaming algorithms for submodular maximization problems have received much attention
recently~\cite{DBLP:conf/icalp/AlalufEFNS20,DBLP:conf/icml/GomesK10,DBLP:conf/kdd/BadanidiyuruMKK14,DBLP:conf/icml/Norouzi-FardTMZ18,DBLP:conf/icml/0001MZLK19}.
Typically, these streaming algorithms require only one or a few passes over the dataset,
store a small portion of items in memory, and compute a solution more efficiently
than the \textsc{Greedy} algorithm at the expense of slightly lower quality.

Despite the extensive studies on streaming submodular maximization, unfortunately, it seems that
none of the existing methods consider the \emph{fairness} issue of the subsets extracted from data streams.
In fact, recent studies~\cite{DBLP:journals/pacmhci/DashSBGGC19,DBLP:journals/cacm/ChouldechovaR20,DBLP:conf/chi/KayMM15,DBLP:conf/icml/CelisKS0KV18}
reveal that data summaries automatically generated by algorithms might be biased with respect to
sensitive attributes such as gender, race, or ethnicity, and the biases in summaries could be passed to
data-driven decision-making processes in education, recruitment, banking, and judiciary systems.
Thus, it is necessary to introduce \emph{fairness} constraints into submodular maximization problems
so that the selected subset can fairly represent each sensitive attribute in the dataset.
Towards this end, we consider that a data stream $V$ comprises $l$ disjoint groups $V_1,V_2,\ldots,V_l$
defined by some sensitive attribute. For example, the groups may correspond to a demographic attribute
such as \emph{gender} or \emph{race}. We define the fairness constraint by assigning a cardinality
constraint $ k_i $ to \textit{each group} $V_i$ and ensuring that $ \sum_{i=1}^{l} k_i = k $.
Then, our goal is to maximize the submodular objective function under the constraint that
the selected subset must contain $k_i$ items from $V_i$. The fairness constraint as defined above
can incorporate different concepts of \emph{fairness} by assigning different values of $ k_1,k_2,\ldots,k_l $.
For example, one can extract a subset that approximately represents the proportion of each group in the dataset
by setting $ k_i=\frac{|V_i|}{|V|} \cdot k $.
As another example, one can enforce a balanced representation of each group by setting $ k_i=\frac{k}{l} $.

Theoretically, the above-defined fairness constraint is a case of \emph{partition matroid}
constraints~\cite{DBLP:conf/kdd/AbbassiMT13,DBLP:conf/icml/JonesNN20,DBLP:conf/icml/KleindessnerAM19},
and thus the optimization problem can be reduced to maximizing submodular set functions with matroid constraints.
It is not surprising that all existing algorithms for submodular maximization with cardinality constraints
cannot be directly used for this problem anymore, because their solutions may not satisfy the fairness constraint
(i.e., the group-specific cardinality constraints).
Nevertheless, a seminal work of Fisher et al.~\cite{Fisher1978} indicates that the \textsc{Greedy} algorithm
with minor modifications is $\frac{1}{2}$-approximate for this problem.
But it still suffers from efficiency issues in the streaming setting.
In addition, the state-of-the-art streaming algorithms~\cite{DBLP:conf/icalp/ChekuriGQ15,DBLP:journals/mp/ChakrabartiK15,DBLP:conf/nips/FeldmanK018}
for submodular maximization with matroid constraints are only $\frac{1}{4}$-approximate
and do not provide solutions of the same quality as the \textsc{Greedy} algorithm efficiently in practice.

In this paper, we investigate the problem of \emph{streaming submodular maximization with fairness constraints}.
Our main contributions are summarized as follows.
\begin{itemize}
  \item We first formally define the \emph{fair submodular maximization} (FSM) problem and show its NP-hardness.
  We also describe the $\frac{1}{2}$-approximation \textsc{Greedy} algorithm for the FSM problem and discuss
  why it cannot work efficiently in data streams. (Section~\ref{sec:def})
  \item We propose a multi-pass streaming algorithm MP-FSM for the FSM problem. Theoretically, MP-FSM
  requires $ O(\frac{1}{\varepsilon}\log\frac{k}{\varepsilon}) $ passes over the dataset,
  stores $O(k)$ items in memory, and has an approximation ratio of $ (\frac{1}{2}-\varepsilon) $
  for any constant $ \varepsilon > 0 $. (Section~\ref{subsec:alg:multi:pass})
  \item We further propose a single-pass streaming algorithm SP-FSM for the FSM problem, which requires
  only one pass over the data stream and offers the same approximation ratio as MP-FSM when an unbounded buffer size
  is permitted. We also discuss how to adapt SP-FSM heuristically to limit the buffer size to $ O(k) $.
  (Sections~\ref{subsec:alg:one:pass} \&~\ref{sec:small_buffer})
  \item Finally, we evaluate the performance of our proposed algorithms against the state-of-the-art methods
  in two real-world application scenarios, namely \emph{maximum coverage on large graphs}
  and \emph{personalized recommendation}. The empirical results on several real-world and synthetic datasets
  demonstrate the efficiency, effectiveness, and scalability of our proposed algorithms. (Section~\ref{sec:exp})
\end{itemize}

%!TEX root = main.tex
\section{Related Work}
\label{sec:literature}

There has been a large body of work on submodular optimization for its wide applications in various real-world problems,
including influence maximization~\cite{DBLP:conf/kdd/KempeKT03,DBLP:journals/pvldb/WangFLT17},
facility location~\cite{DBLP:conf/kdd/LeskovecKGFVG07,DBLP:conf/nips/LindgrenWD16},
nonparametric learning~\cite{DBLP:conf/icml/GomesK10,DBLP:conf/kdd/BadanidiyuruMKK14},
and group recommendation~\cite{DBLP:conf/www/SerbosQMPT17}.
We refer interested readers to~\cite{DBLP:books/cu/p/0001G14} for a survey.

The line of research that is the most relevant to this work is \emph{streaming algorithms for submodular maximization}.
The seminal work of Fisher, Nemhauser, and Wolsey~\cite{DBLP:journals/mp/NemhauserWF78,Fisher1978} showed that
the \textsc{Greedy} algorithm, which iteratively added an item that maximally increased the utility with $k$ passes
over the dataset, gave approximation ratios of $ (1-\frac{1}{e}) $ and $ \frac{1}{2} $ for maximizing monotone
submodular functions with cardinality and matroid constraints, respectively.
Then, a series of recent studies~\cite{DBLP:conf/icml/GomesK10,DBLP:conf/kdd/BadanidiyuruMKK14,DBLP:journals/topc/KumarMVV15,DBLP:conf/icml/0001MZLK19}
proposed multi- or single-pass streaming algorithms for maximizing monotone submodular functions
subject to cardinality constraints with the same approximation ratio of $ (\frac{1}{2}-\varepsilon) $.
Furthermore, Norouzi-Fard et al.~\cite{DBLP:conf/icml/Norouzi-FardTMZ18} showed that any single-pass streaming algorithm
must use $ \Omega(\frac{n}{k}) $ memory to achieve an approximation ratio of over $ \frac{1}{2} $.
They also proposed streaming algorithms with approximation factors better than $ \frac{1}{2} $
by assuming that items arrived in random order or running in multiple passes.
Alaluf et al.~\cite{DBLP:conf/icalp/AlalufEFNS20} proposed a $0.2779$-approximation streaming algorithm
for maximizing non-monotone submodular functions with cardinality constraints.
Moreover, streaming submodular maximization is also studied in different models,
e.g., the sliding-window model~\cite{DBLP:conf/www/EpastoLVZ17,DBLP:journals/tkde/WangLT19}
where only recent items within a time window are available for selection,
the time-decay model~\cite{DBLP:conf/aaai/ZhaoSWLZ19} where the weights of old items decrease over time,
and the deletion-robust model~\cite{DBLP:conf/icml/MirzasoleimanK017,DBLP:conf/nips/MitrovicBNTC17,DBLP:conf/icml/0001ZK18}
where existing items might be removed from the stream.
However, all above steaming algorithms are specific for the cardinality constraint
and cannot be directly used for the fairness constraint in this paper.
We note that Kazemi et al.~\cite{DBLP:conf/icml/0001ZK18} also introduce \emph{fairness}
into submodular maximization problems.
However, they consider removing sensitive items from the dataset for
ensuring fairness, which is different from the problem we study in this paper.

Chakrabarti and Kale~\cite{DBLP:journals/mp/ChakrabartiK15} proposed a $\frac{1}{4p}$-approximation single-pass
streaming algorithms for maximizing monotone submodular functions with the intersections of $p$ matroid constraints.
Chekuri et al.~\cite{DBLP:conf/icalp/ChekuriGQ15} generalized the algorithm in~\cite{DBLP:journals/mp/ChakrabartiK15}
to the case of non-monotone submodular functions. Both algorithms achieve a $\frac{1}{4}$-approximation ratio for
the FSM problem. Chan et al.~\cite{DBLP:conf/soda/Chan0JKT17} improved the approximation ratio for partition matroids
to $0.3178$ via randomization and relaxation. Feldman et al.~\cite{DBLP:conf/nips/FeldmanK018} introduced a subsampling
method to speed up the algorithm of~\cite{DBLP:conf/icalp/ChekuriGQ15}
while still achieving an approximation ratio of $\frac{1}{4p}$ in expectation.
Huang et al.~\cite{DBLP:conf/approx/HuangTW20} proposed an $ O(\frac{1}{\varepsilon}) $-pass
$ \frac{1}{2+\varepsilon} $-approximation algorithm for monotone submodular maximization with matroid constraints.
We implement the aforementioned algorithms from~\cite{DBLP:conf/icalp/ChekuriGQ15,DBLP:journals/mp/ChakrabartiK15,DBLP:conf/nips/FeldmanK018,DBLP:conf/approx/HuangTW20} as baselines in our experiments.
We do not implement the algorithm in~\cite{DBLP:conf/soda/Chan0JKT17}
since it is not scalable to large datasets.

Another line of research related to this work is \emph{fair data summarization}.
Fair $k$-center for data summarization was studied
in~\cite{DBLP:conf/icml/KleindessnerAM19,DBLP:conf/icml/JonesNN20,chiplunkar2020solve}.
Celis et al.~\cite{DBLP:conf/icml/CelisKS0KV18} proposed a determinantal point process (DPP) based
sampling method for fair data summarization.
Dash et al.~\cite{DBLP:journals/pacmhci/DashSBGGC19} considered the fairness issue on summarizing
user-generated textual content.
Although these studies adopt similar definitions of fairness constraints to ours, their proposed methods cannot be
applied to the FSM problem since the objective functions of the problems they study are not submodular.

%!TEX root = main.tex
\section{Problem Definition}\label{sec:def}

We consider the problem of selecting a subset of representative items from a dataset $V$ of size $n$.
Our goal is to maximize a non-negative set function $ f : 2^{V} \rightarrow \mathbb{R}^{+} $, where,
for any subset $ S \subseteq V $, $ f(S) $ quantifies the utility of $S$, i.e., how well $S$ represents $V$
according to some objective. In many data summarization problems
(e.g.,~\cite{DBLP:conf/icml/GomesK10,DBLP:conf/kdd/BadanidiyuruMKK14,DBLP:conf/nips/LindgrenWD16,DBLP:conf/www/EpastoLVZ17}),
the utility function satisfies an intuitive \emph{diminishing returns} property called \emph{submodularity}.
To describe it formally, we define the \emph{marginal gain} $ \Delta_f(v|S) \coloneqq f(S\cup\{v\})-f(S) $
as the increase in utility when an item $v$ is added to a set $S$.
A set function $f$ is \emph{submodular} iff $ \Delta_f(v|A) \geq \Delta_f(v|B) $
for any $ A \subseteq B \subseteq V $ and $ v \in V \setminus B $.
This means that adding an item $e$ to a set $A$ leads to at least as much utility gain as adding $v$
to a superset $B$ of $A$. Additionally, a submodular function $f$ is \emph{monotone} iff $ \Delta_f(v|S) \geq 0 $
for any $ S \subseteq V $ and $ v \in V \setminus S $, i.e., adding any new item $v$ never decreases the utility of $S$.
In this work, we assume that the function $f$ is both monotone and submodular.
Moreover, following most existing works~\cite{DBLP:conf/icml/GomesK10,DBLP:conf/kdd/BadanidiyuruMKK14,DBLP:conf/kdd/LeskovecKGFVG07,DBLP:conf/www/EpastoLVZ17,DBLP:conf/icalp/ChekuriGQ15,DBLP:conf/nips/FeldmanK018,DBLP:conf/icml/0001MZLK19,DBLP:conf/icml/Norouzi-FardTMZ18}, we assume that the utility $ f(S) $ of any set $ S \subseteq V $ is given by a value oracle -- i.e., the value of $ f(S) $ is retrieved in constant time.

Let us consider the following canonical optimization problem: given a monotone submodular set function $f$ and a dataset $V$, find a subset of size $k$ from $V$ that maximizes the function $f$, i.e.,
\begin{equation}\label{def:standard}
  \max_{S \subseteq V} \; f(S) \quad \textnormal{s.t.} \quad |S| = k
\end{equation}
The problem in Eq.~\ref{def:standard} is referred to as the cardinality-constrained submodular maximization (CSM) problem and proven to be NP-hard~\cite{DBLP:journals/jacm/Feige98} for many classes of submodular functions.
And the well-known greedy algorithm of Nemhauser et al.~\cite{DBLP:journals/mp/NemhauserWF78} achieves a $(1-\frac{1}{e})$-approximation for this problem.

Now we introduce the \emph{fairness} issue into the CSM problem. Let $ [l] = \{1,\ldots,l\} $.
Suppose that the dataset $V$ is partitioned into $l$ (disjoint) groups, each of which corresponds to a sensitive class, and $V_i$ is the set of items from the $i$-th group in $V$ with $ \bigcup_{i=1}^{l} V_i = V $.
Then, for each group, we demand that the solution $S$ must contain $k_i$ items from $V_i$, and $ \sum_{i=1}^{l} k_i=k $.
Formally, the fair submodular maximization (FSM) problem is defined as follows:
\begin{equation}\label{def:fsm}
  S^* = \argmax_{S \subseteq V} \; f(S) \quad \textnormal{s.t.} \quad |S \cap V_i| = k_i, \forall i \in [l]
\end{equation}
where $S^*$ and $\mathtt{OPT}=f(S^*)$ denote the optimal solution and its utility.
The values of $ k_1,\ldots,k_l \in \mathbb{Z}^{+} $ are given as input to the problem (here, we assume $k_i > 0 $ since we can simply ignore all items in $V_i$ if $ k_i=0 $) and determined according to the notion of fairness.
For example, one can use $ k_i = \frac{n_i}{n} \cdot k $ where $ n_i = |V_i| $ to obtain a \emph{proportional representation}.
As another example, an \emph{equal representation} can be acquired by setting $ k_i=\frac{k}{l} $ for all $ i \in [l] $.

\begin{algorithm}[t]
  \caption{\textsc{Greedy}}\label{alg:greedy}
  \Input{Dataset $ V $, groups $ V_1,\ldots,V_l \subseteq V $, total size constraint $ k \in \mathbb{Z}^{+} $,
         group size constraints $ k_1,\ldots,k_l \in \mathbb{Z}^{+} $}
  \Output{Solution $S$ for the FSM problem on $V$}
  Initialize the solution $S \gets \varnothing$\;
  \For{$j \gets 1,\ldots,k$}{
    \For{$i \gets 1,\ldots,l$}{
      \uIf{$|S \cap V_i| < k_i$}{
        $v^*_i \gets \argmax_{v \in V_i} \Delta_f(v|S)$\;
      }
      \Else{
        $v^*_i \gets \textnormal{NULL}$\;
      }
    }
    $v^* \gets \argmax_{i \in [l] \,:\, v^*_i \neq \textnormal{NULL}} \Delta_f(v^*_i|S)$\;
    $S \gets S \cup \{v^*\}$, $V \gets V \setminus \{v^*\}$\;
  }
  \Return{$S$}\;
\end{algorithm}

The FSM problem in Eq.~\ref{def:fsm} is still NP-hard because the CSM problem in Eq.~\ref{def:standard}
is its special case when $l=1$.
Nevertheless, a modified \textsc{Greedy} algorithm first proposed in~\cite{Fisher1978} provides a $\frac{1}{2}$-approximate solution for the FSM problem, since the fairness constraint we consider is a case of the \emph{partition matroid} constraint.
The procedure of \textsc{Greedy} is described in Algorithm~\ref{alg:greedy}.
Starting from $S=\varnothing$, it iteratively adds an item $v^*$ with the maximum utility gain $\Delta_f(v^*|S)$ to the current solution $S$.
To guarantee that solution $S$ satisfies the fairness constraint, it excludes from consideration all items of $V_i$ once there are $k_i$ items from $V_i$ in $S$, i.e., $ |S \cap V_i|=k_i $.
The solution $S$ after $k$ iterations is returned for the FSM problem.
The running time of \textsc{Greedy} is $O(nk)$ because it runs $k$ passes through the dataset and evaluates the value of $f$ at most $n$ times per pass for identifying $v^*_i$.
Therefore, \textsc{Greedy} becomes very inefficient when the dataset size is large; even worse, \textsc{Greedy} cannot work in the single-pass streaming setting if the dataset does not fit in the memory.
In what follows, we will investigate the FSM problem in streaming settings.

%!TEX root = main.tex
\section{Our Algorithms}
\label{sec:alg}

In this section, we present our proposed algorithms for the fair submodular maximization (FSM) problem in data streams.
Firstly, we propose a multi-pass streaming algorithm called MP-FSM.
For any constant $\varepsilon \in (0,1)$, MP-FSM requires $ O\big(\frac{1}{\varepsilon}\log\frac{k}{\varepsilon}\big) $ passes over the dataset, stores $O(k)$ items in memory, and provides a $ \frac{1}{2}(1-\varepsilon) $-approximate solution for the FSM problem.
Secondly, we propose a single-pass streaming algorithm called SP-FSM on the top of MP-FSM.
SP-FSM has an approximation ratio of $(\frac{1}{2}-\varepsilon)$ and sublinear update time per item.
But it might keep $O(n)$ items in a buffer for post-processing in the worst case, and thus its space complexity is $O(n)$.
Therefore, we further discuss how to restrict the buffer size of SP-FSM when the memory space is limited and how the approximation ratio of SP-FSM is affected accordingly.

\subsection{Multi-Pass Streaming Algorithm}
\label{subsec:alg:multi:pass}

In this subsection, we present our multi-pass streaming algorithm called MP-FSM for the FSM problem.
In general, MP-FSM adopts a threshold-based approach similar to existing streaming algorithms for the CSM
problem~\cite{DBLP:conf/kdd/BadanidiyuruMKK14,DBLP:journals/topc/KumarMVV15,DBLP:conf/icml/Norouzi-FardTMZ18,DBLP:conf/icml/0001MZLK19}.
The high-level idea of the threshold-based approach is to process items in a data stream sequentially with a threshold $\tau$: for each item $v$ received from the stream, it will accept $v$ into a solution $S$ if $\Delta_f(v|S)$ reaches $\tau$ and discard $v$ otherwise.
But differently from most thresholding algorithms~\cite{DBLP:conf/kdd/BadanidiyuruMKK14,DBLP:journals/topc/KumarMVV15,DBLP:conf/icml/0001MZLK19} for the CSM problem, which run in only one pass and use a fixed threshold for each candidate solution, MP-FSM scans the dataset in multiple passes using a decreasing threshold to determine whether to include an item in each pass so that the solution has a constant approximation ratio while satisfying the fairness constraint.

We present the detailed procedure of MP-FSM in Algorithm~\ref{alg:mp:fsm}.
In the first pass, it finds the item $v_{max}$ with the maximum utility $ \delta_{max}=f(\{v_{max}\}) $ among all items in the dataset $V$.
The purpose of finding $v_{max}$ is to determine the range of thresholds to be used in subsequent passes.
Meanwhile, it keeps a random sample $R_i$ of $k_i$ items uniformly from $V_i$ for each $i \in [l]$, which will be used for post-processing to guarantee that the solution satisfies the fairness constraint.
Then, it initializes a solution $S$ containing only $v_{max}$ and a threshold $\tau = (1-\varepsilon)\cdot\delta_{max}$ for the second pass.
After that, it scans the dataset $V$ sequentially in multiple passes.
In each pass, it decreases the threshold $\tau$ by $(1-\varepsilon)$ times and adds an item $v \in V_i$ to the current solution $S$ if the marginal gain of $v$ w.r.t.~$S$ reaches $\tau$ and there are fewer than $k_i$ items in $S$ from $V_i$.
When the solution $S$ has contained $k$ items or the threshold $\tau$ has been decreased to be lower than $\frac{\varepsilon}{k} \cdot \delta_{max}$, no more passes are needed.
Finally, if the solution $S$ does not satisfy the fairness constraint, it will add items from random samples to $S$ for ensuring its validity.

\begin{algorithm}[t]
  \caption{MP-FSM}\label{alg:mp:fsm}
  \Input{Dataset $V$, groups $V_1,\ldots,V_l \subseteq V$, total size constraint $k \in \mathbb{Z}^{+}$,
         group size constraints $k_1,\ldots,k_l \in \mathbb{Z}^{+}$, parameter $ \varepsilon \in (0,1)$}
  \Output{Solution $S$ for the FSM problem on $V$}
  \tcc{Pass 1: Get $v_{max}$ and reservoir sampling}
  $v_{max} \gets \argmax_{v \in V} f(\{v\})$ and $ \delta_{max} \gets f(\{v_{max}\})$\;
  Keep a random sample $R_i$ of $k_i$ items uniformly from $V_i$ for each $i\in[l]$
  via reservoir sampling~\cite{DBLP:journals/toms/Vitter85}\;
  \tcc{Pass $2$ to $p$: Compute solution $S$}
  $ S \gets \{v_{max}\} $ and $ \tau \gets (1-\varepsilon)\cdot\delta_{max} $\;
  \While{$ \tau > \frac{\varepsilon}{k} \cdot \delta_{max} $}{
    \ForEach{item $ v \in V \setminus S $}{
      \If{$ v \in V_i $ and $ |S \cap V_i| < k_i $ and $ \Delta_f(v|S) \geq \tau $}{
        $ S \gets S \cup \{v\} $\;
      }
    }
    \uIf{$ |S|=k $}{
      \textbf{break}\;
    }
    \Else{
      $ \tau \gets (1-\varepsilon) \cdot \tau $\;
    }
  }
  \tcc{Post processing: Ensure fairness}
  \While{$ \exists i \in [l] : |S \cap V_i| < k_i $}{
    Add items in $R_i$ to $S$ until $|S \cap V_i| = k_i$\;
  }
  \Return{$S$}\;
\end{algorithm}

Next, we provide some theoretical analyses for the MP-FSM algorithm.
First, we give the approximation ratio of MP-FSM in Theorem~\ref{thm:approx:mp:fsm}. And then, the complexity of MP-FSM
is analyzed in Theorem~\ref{thm:complexity:mp:fsm}.

\begin{theorem}\label{thm:approx:mp:fsm}
  For any parameter $\varepsilon \in (0,1)$, MP-FSM in Algorithm~\ref{alg:mp:fsm} is a $\frac{1}{2}(1-\varepsilon)$-approximation algorithm for the FSM problem.
\end{theorem}
\begin{proof}
  Let $O$ be the optimal solution for the FSM problem on dataset $V$ and $O_i = O \cap V_i$ be the intersection of $O$ and $V_i$ for each $i \in [l]$.
  We consider that MP-FSM runs in $p$ passes and $S^{(j)}$ ($ 1 \leq j \leq p $) is the partial solution
  of MP-FSM after $j$ passes.
  For any subset $O_i$ of $O$ and the solution $S^{(p)}$ after $p$ passes, we have either (1) $ |S^{(p)} \cap V_i| = k_i $ or (2) $ |S^{(p)} \cap V_i| < k_i $.
  If $ |S^{(p)} \cap V_i| = k_i $, there are two cases for each item $o \in O_i$: (1.1) $o \in S^{(p)}$ and (1.2) $o \notin S^{(p)}$.
  In Case (1.1), we have $ \Delta_f(o|S^{(p)}) = 0 $.
  In Case (1.2), we compare $o$ with an item $s$ from $V_i$ added to the solution during the $j$-th pass.
  Since both $o$ and $s$ cannot be added in the $(j-1)$-th pass and $ |S^{(j-1)} \cap V_i| < k_i $, it is safe to say that the marginal gains of $o$ and $s$ w.r.t.~$S^{(j-1)}$ do not reach the threshold $\tau^{(j-1)}$ of the $(j-1)$-th pass.
  As $s$ is added in the $j$-th pass, we have $\Delta_f(s|S^{\prime}) \geq \tau^{(j)}$ where $S^{\prime} \subseteq S^{(j)}$ is the partial solution before $s$ is added.
  Therefore, we have the following sequence of inequalities:
  \begin{equation}\label{eq:case:1.2}
    \Delta_f(o|S^{(p)}) \leq \Delta_f(o|S^{(j-1)}) < \tau^{(j-1)} = \frac{\tau^{(j)}}{1-\varepsilon}
    \leq \frac{\Delta_f(s|S^\prime)}{1-\varepsilon}
  \end{equation}
  Then, if $ |S^{(p)} \cap V_i| < k_i $, there are also two cases for $o \in O_i$: (2.1) $o \in S^{(p)}$ and (2.2) $o \notin S^{(p)}$.
  Case (2.1) is exactly the same as Case (1.1).
  In Case (2.2), we have:
  \begin{equation}\label{eq:case:2.2}
    \Delta_f(o|S^{(p)}) < \tau^{(p)}
    \leq \frac{\varepsilon}{k(1-\varepsilon)} \cdot \delta_{max}
  \end{equation}
  where $\tau^{(p)}$ is the threshold of the $p$-th pass.

  Next, we divide $O$ into two disjoint subsets $O^{\prime}$ and $O^{\prime\prime}$ as follows: $ O^{\prime} = \bigcup_{i^{\prime}} O_{i^{\prime}} $ where $ |S^{(p)} \cap V_{i^{\prime}}| = k_{i^{\prime}} $, i.e., all items from groups satisfying Case (1), and $O^{\prime\prime}=O \setminus O^{\prime}$, i.e., all items from groups satisfying Case (2).
  We define an injection $ \pi: O^{\prime} \rightarrow S^{(p)} $ that maps each item in $O^{\prime}$ to an item in $S^{(p)}$ as follows: If $ o \in S^{(p)} $, then $ \pi(o) = o $; otherwise, $ \pi(o) $ will be an arbitrary item $ s \in S^{(p)} $ from the same group as $o$ and $ s \notin O $.
  Based on the result of Eq.~\ref{eq:case:1.2}, we can get the following inequalities for $O^{\prime}$:
  \begin{equation}\label{eq:case:1}
    \sum_{o \in O^{\prime}} \Delta_f(o|S^{(p)})
    \leq \frac{\sum_{\pi(o) \in S^{(p)}}\Delta_f(\pi(o)|S^\prime)}{1-\varepsilon}
    \leq \frac{f(S^{(p)})}{1-\varepsilon}
  \end{equation}
  Here, $S^\prime$ denotes the partial solution before $\pi(o)$ is added and the second inequality is acquired from the fact that $ f(S^{(p)})=\sum_{s \in S^{(p)}} \Delta_f(s|S^\prime) $.
  Then, based on the result of Eq.~\ref{eq:case:2.2}, we have the following inequalities for $O^{\prime\prime}$:
  \begin{equation}\label{eq:case:2}
    \sum_{o \in O^{\prime\prime}} \Delta_f(o|S^{(p)})
    \leq \frac{\varepsilon |O^{\prime\prime}|}{k(1-\varepsilon)} \cdot \delta_{max}
    \leq \frac{\varepsilon}{1-\varepsilon} \cdot f(S^{(p)})
  \end{equation}
  because $ |O^{\prime\prime}| < k $ and $ \delta_{max} \leq f(S^{(p)}) $.
  Finally, we have the following sequence of inequalities from Eq.~\ref{eq:case:1} and~\ref{eq:case:2}:
  \begin{multline*}
    f(O \cup S^{(p)}) - f(S^{(p)})
    = \sum_{o \in O^{\prime}}\Delta_f(o|S^{(p)}) + \sum_{o \in O^{\prime\prime}}\Delta_f(o|S^{(p)}) \\
    \leq \frac{1}{1-\varepsilon} \cdot f(S^{(p)})+\frac{\varepsilon}{1-\varepsilon} \cdot f(S^{(p)})
    = \frac{1+\varepsilon}{1-\varepsilon} \cdot f(S^{(p)})
  \end{multline*}
  Since $ \mathtt{OPT} = f(O) \leq f(O \cup S^{(p)}) $,
  we get $ \mathtt{OPT} \leq (1 + \frac{1+\varepsilon}{1-\varepsilon}) \cdot f(S^{(p)}) \leq \frac{2}{1-\varepsilon} \cdot f(S^{(p)}) $.
  Finally, we conclude the proof from the fact that $ f(S) \geq f(S^{(p)}) \geq \frac{1}{2}(1-\varepsilon) \mathtt{OPT} $.
\end{proof}

\begin{theorem}\label{thm:complexity:mp:fsm}
  MP-FSM in Algorithm~\ref{alg:mp:fsm} requires $ O\big(\frac{1}{\varepsilon}\log\frac{k}{\varepsilon}\big) $ passes over the dataset $V$, stores at most $O(k)$ items,
  and has $ O\big(\frac{n}{\varepsilon}\log\frac{k}{\varepsilon}\big) $ time complexity.
\end{theorem}
\begin{proof}
  First of all, since the threshold $\tau$ is decreased by $(1-\varepsilon)$ times after one pass, $ \tau^{(2)}=(1-\varepsilon)\cdot\delta_{max} $, and $ \tau^{(p)} \geq \frac{\varepsilon}{k} \cdot \delta_{max}$, we get $ (1-\varepsilon)^{p-1} \geq \frac{\varepsilon}{k} $.
  Taking the logarithm on both sides of the last inequality and the Taylor expansion of $\log(1-\varepsilon)$,
  we have $ p-1 \leq \frac{1}{\log(1-\varepsilon)} \cdot \log\frac{\varepsilon}{k}
  \leq \frac{1}{\varepsilon} \log\frac{k}{\varepsilon} $ and thus the number $p$ of passes in MP-FSM is $ O\big(\frac{1}{\varepsilon}\log\frac{k}{\varepsilon}\big) $.
  Furthermore, MP-FSM only stores items in the solution and random samples for post-processing, both of which contain at most $k$ items.
  Hence, MP-FSM stores at most $O(k)$ items.
  Finally, because MP-FSM evaluates the value of function $f$ at most $n$ times per pass, the total number of function evaluations in MP-FSM is $ O\big(\frac{n}{\varepsilon}\log\frac{k}{\varepsilon}\big) $.
\end{proof}

\subsection{Single-Pass Streaming Algorithm}\label{subsec:alg:one:pass}

In this subsection, we present our single-pass streaming algorithm called SP-FSM for the FSM problem.
Generally, SP-FSM is based on a threshold-based approach, similar to MP-FSM.
However, several adaptations are required so that SP-FSM can provide an approximate solution in only one pass over the dataset.
First of all, because $ v_{max} $ and $ \delta_{max} $ are unknown in advance, SP-FSM should keep track of them from received items, dynamically decide a sequence of thresholds based on the observed $ \delta_{max} $, and maintain a candidate solution for each threshold (instead of keeping only one solution over multiple passes in MP-FSM).
Furthermore, as only one pass is permitted, an item will be unrecoverable once it is discarded.
To provide a theoretical guarantee for the quality of solutions in adversarial settings, SP-FSM keeps a buffer to store items that are neither included into solutions nor safely discarded.
Finally, whenever a solution is requested during the stream, SP-FSM will reconsider the buffered items for post-processing by attempting to add them greedily to candidate solutions.
We will show that SP-FSM has an approximation ratio of $(\frac{1}{2}-\varepsilon)$ with a judicious choice of parameters when the buffer size is unlimited.

The detailed procedure of SP-FSM is presented in Algorithm~\ref{alg:op:fsm}.
Here, $ \delta_{max} $ keeps the maximum utility of any single item among all items received so far, $\mathtt{LB}$ maintains the lower bound of $\mathtt{OPT}$ estimated from candidate solutions, $B$ stores the buffered items, and $R_i$ is a set of $k_i$ items sampled uniformly from all received items in $V_i$.
In addition, two parameters $\alpha$ and $\beta$ affect the number of candidate solutions and the number of buffered items, respectively.
For larger values of $\alpha$, the gaps between neighboring thresholds are bigger and thus the numbers of candidate solutions are fewer;
for larger values of $\beta$, the conditions for adding an item to the buffer are more rigorous and naturally the buffer sizes are smaller.
The procedure for stream processing of SP-FSM is given in Line~\ref{line:op:stream:s}--\ref{line:op:stream:t}.
For each item $v \in V_i$ received from $V$, it first updates the value of $ \delta_{max} $ and the sample $R_i$ w.r.t.~$v$.
Then, it maintains a sequence $T$ of thresholds picked from a geometric progression $ \{(1+\alpha)^{j} | j \in \mathbb{Z}\} $ and a candidate solution $ S_{\tau} $ for each $ \tau \in T $.
Specifically, the upper bound of the threshold is set to $ \delta_{max} $ since $S_{\tau}=\varnothing$
for any $ \tau > \delta_{max} $; the lower bound is set to $\frac{\max\{\delta_{max},\mathtt{LB}\}}{2k}$ because any candidate with a threshold lower than $\frac{\mathtt{OPT}}{2k}$ is safe to be discarded (as shown in our theoretical analysis later) and $\max\{\delta_{max},\mathtt{LB}\}$ is the lower bound of $\mathtt{OPT}$.
After maintaining the thresholds and their corresponding candidates, SP-FSM evaluates the marginal gain $\Delta_f(v|S_{\tau})$ of $v$ for each candidate $ S_{\tau} $ with threshold $ \tau \in T $. Similar to MP-FSM, it will add $v$ to $ S_{\tau} $ if $\Delta_f(v|S_{\tau})$ reaches $\tau$ and $|S_{\tau} \cap V_i| < k_i$. Additionally, it will add $v$ to the buffer $B$ if $\Delta_f(v|S_{\tau})$ is at least $\frac{\beta\cdot \mathtt{LB}}{k}$ but less than $\tau$.
Finally, $\mathtt{LB}$ is updated to the utility of the best solution found so far.
The procedure for post-processing of SP-FSM is shown in Lines~\ref{line:op:post:s}--\ref{line:op:post:t}.
It first finds out the smallest $ \tau \in T $ such that $ |S_{\tau} \cap V_i| < k_i $ for each $i \in [l]$ as $\tau^{\prime}$; if such $ \tau $ does not exist, i.e., there exists some $i$ such that $ |S_{\tau} \cap V_i| = k_i $ for every $S_{\tau}$, the largest $ \tau \in T $ is used as $\tau^{\prime}$.
For each $ \tau \leq \tau^{\prime}$ in $ T $, it runs \textsc{Greedy} in Algorithm~\ref{alg:greedy} to reevaluate the items in $B$ and $R_i$ and add them to $S_{\tau}$ until $ |S_{\tau}|=k $.
Lastly, the candidate solution with the maximum utility after post-processing is returned as the final solution.

\begin{algorithm}[t]
  \caption{SP-FSM}\label{alg:op:fsm}
  \Input{Data stream $V$, groups $V_1,\ldots,V_l \subseteq V$, total size constraint $k \in \mathbb{Z}^{+}$,
         group size constraints $k_1,\ldots,k_l \in \mathbb{Z}^{+}$,
         parameters $ \alpha,\beta \in (0,1)$}
  \Output{Solution $S$ for the FSM problem on $V$}
  $ \delta_{max} \gets 0 $, $\mathtt{LB} \gets 0$, $ B \gets \varnothing $, and $ R_i \gets \varnothing $ for each $ i \in [l] $\;
  \tcc{Stream processing}
  \ForEach{item $v \in V_i$ received from $V$ \label{line:op:stream:s}}{
    $ \delta_{max} \gets \max\{ \delta_{max},f(\{v\}) \} $\;
    Update $R_i$ w.r.t.~$v$ using reservoir sampling~\cite{DBLP:journals/toms/Vitter85}\;
    $ T \gets \{(1+\alpha)^{j} | j \in \mathbb{Z}, \frac{\max\{\delta_{max},\mathtt{LB}\}}{2k} \leq (1+\alpha)^{j} \leq \delta_{max} \} $\;\label{line:op:range}
    Discard $ S_{\tau} $ for all $ \tau \notin T $\;
    Initialize $ S_{\tau} \gets \varnothing $ for each $\tau$ newly added to $T$\;
    \ForEach{$ \tau \in T $}{
      \If{$|S_{\tau} \cap V_i| < k_i$}{
        \uIf{$\Delta_f(v|S_{\tau}) \geq \tau$}{
          $ S_{\tau} \gets S_{\tau} \cup \{v\} $\;
        }
        \ElseIf{$\Delta_f(v|S_{\tau}) \geq \frac{\beta \cdot \mathtt{LB}}{k}$}{
          $ B \gets B \cup \{v\} $\;
        }
      }
    }
    $\mathtt{LB} \gets \max_{\tau \in T} f(S_{\tau})$\;\label{line:op:stream:t}
  }
  \tcc{Post processing}
  Let $\tau^{\prime}$ be the smallest $ \tau \in T $ such that $ |S_{\tau} \cap V_i| < k_i $ for each $i \in [l]$
  or the largest $ \tau \in T $ if there exists some $i$ such that $ |S_{\tau} \cap V_i| = k_i $ for every $S_{\tau}$\;
  \label{line:op:post:s}
  \ForEach{$ \tau \leq \tau^{\prime}$ in $T$}{
    Run \textsc{Greedy} in Algorithm~\ref{alg:greedy} to add items from buffer $B$ and samples $R_i$ for all $i \in [l]$
    to $S_{\tau}$ until $ |S_{\tau}|=k $\;\label{line:op:post:t}
  }
  \Return{$S \gets \argmax_{\tau \in T} f(S_{\tau})$}\;
\end{algorithm}

Next, we will provide the theoretical analyses for the SP-FSM algorithm.
First, in Lemma~\ref{lm:approx:know:opt}, we analyze the special cases when the solution returned after stream processing (without post-processing) can achieve a good approximation ratio.

\begin{lemma}\label{lm:approx:know:opt}
  Assume that $ \frac{\mathtt{OPT}}{2k} \leq \tau \leq \frac{(1+\alpha)\cdot\mathtt{OPT}}{2k} $. If either $ |S_{\tau}| = k $ or $ |S_{\tau} \cap V_i| < k_i $ for all $i \in [l]$, then $ f(S_{\tau}) \geq \frac{1-\alpha}{2} \cdot \mathtt{OPT} $.
\end{lemma}
\begin{proof}
  First of all, when $ |S_{\tau}| = k $, it holds that $ f(S_{\tau}) \geq k\tau \geq k \cdot \frac{\mathtt{OPT}}{2k} = \frac{1}{2} \cdot \mathtt{OPT} \geq \frac{1-\alpha}{2} \cdot \mathtt{OPT}$.
  Then, when $ |S_{\tau} \cap V_i| < k_i $ for all $i \in [l]$, we have $\Delta_f(v|S_{\tau}) < \tau$ for any $ v \in V \setminus S_{\tau} $.
  Let $O$ be the optimal solution for the FSM problem on $V$.
  We can acquire that
  \begin{align*}
    f(O \cup S_{\tau}) - f(S_{\tau}) & \leq \sum_{o \in O \setminus S_{\tau}} \Delta_f(o|S_{\tau}) < k \tau \\
    & \leq k \cdot \frac{(1+\alpha)\cdot\mathtt{OPT}}{2k} = (1+\alpha)\cdot\frac{\mathtt{OPT}}{2}
  \end{align*}
  Therefore, we have $ f(S_{\tau}) \geq f(O \cup S_{\tau}) - (1+\alpha)\cdot\frac{\mathtt{OPT}}{2}  \geq \mathtt{OPT} - (1+\alpha)\cdot\frac{\mathtt{OPT}}{2} = \frac{1-\alpha}{2} \cdot \mathtt{OPT} $.
  We conclude the proof by considering both cases collectively.
\end{proof}

Lemma~\ref{lm:approx:know:opt} is useful because one of the thresholds $\tau \in T$ of SP-FSM (Line~\ref{line:op:range} of Algorithm~\ref{alg:op:fsm}) must satisfy the first condition $\frac{\mathtt{OPT}}{2k} \leq \tau \leq \frac{(1+\alpha)\cdot\mathtt{OPT}}{2k}$  of the lemma.
This is because $T$ is a geometric progression with a scale factor of $(1+\alpha)$ and spans the range $[\frac{\max\{\delta_{max},\mathtt{LB}\}}{2k}, \delta_{max}]$, with $ \max\{\delta_{max},\mathtt{LB}\} \leq \mathtt{OPT} \leq k \cdot \delta_{max} $.

This implies that, if the remaining conditions of Lemma~\ref{lm:approx:know:opt} were satisfied as well, the solution of SP-FSM after stream processing would have the strong approximation guarantee given by Lemma~\ref{lm:approx:know:opt}.
Intuitively, this would be the case when the utility distribution of items was generally ``balanced'' among groups, so that either all or none of the group budgets would be exhausted by the end of stream processing.
However, in case that the utilities are highly imbalanced among groups, the approximation ratio would become significantly lower.
On the one hand, SP-FSM might miss high-utility items in some groups from the stream because the threshold is too low and the solution has been filled by earlier items with lower utilities in these groups.
On the other hand, SP-FSM might not include enough items from the other groups because the threshold is too high for them.
Note that, for $ \frac{\mathtt{OPT}}{2k} \leq \tau \leq \frac{(1+\alpha)\cdot\mathtt{OPT}}{2k} $, Lemma~\ref{lm:approx:know:opt} allows the approximation factor of $S_{\tau}$ to drop to $ \frac{\min_{i\in[l]} k_i\tau}{\mathtt{OPT}} \geq \min_{i\in[l]} \frac{k_i}{2k} \geq \frac{1}{2k} $ when some group budgets are exhausted but the others are not.

Therefore, we further include the buffer and post-processing procedures in SP-FSM so that it still achieves a constant approximation independent of $k$ for an arbitrary group size constraint.
In Lemma~\ref{lm:approx:post}, we analyze the approximation ratio of the solution returned by SP-FSM after post-processing.

\begin{lemma}\label{lm:approx:post}
  Let $\tau^{\prime}$ be chosen according to Line~\ref{line:op:post:s} of Algorithm~\ref{alg:op:fsm}. It holds that $ f(S_{\tau^{\prime}}) \geq \frac{1-\beta}{2 + \alpha} \cdot \mathtt{OPT} $ after post-processing.
\end{lemma}
\begin{proof}
  We consider two cases separately: (1)  $|S_{\tau^{\prime}} \cap V_i| < k_i $ for each $i \in [l]$ or (2) $ \tau^{\prime} $ is the maximum in $T$.
  In Case (1), we divide the items in the optimal solution $O$ into three disjoint subsets: $O_1 = O \cap S_{\tau^{\prime}}$, i.e., items included in $ S_{\tau^{\prime}} $ during stream and post processing;  $ O_2 = O \cap (B \setminus S_{\tau^{\prime}})$, i.e., items stored in the buffer but not added to $ S_{\tau^{\prime}} $; $ O_3 = O \cap (V \setminus (B\cup S_{\tau^{\prime}}))$, i.e., items discarded during stream processing.
  For each $ o \in O_2 $, we can always find an item $ s \in S_{\tau^{\prime}} $ from the same group as $o$ such that $ \Delta_f(s|S^{\prime}) \geq \Delta_f(o|S^{\prime}) \geq \Delta_f(o|S_{\tau^{\prime}}) $ where $ S^{\prime} \subseteq S_{\tau^{\prime}} $ is the partial solution when $s$ is added.
  This is because \textsc{Greedy} always picks the item with the maximum marginal gain within each group.
  In addition, for each $ o \in O_3 $, we have $ \Delta_f(o|S_{\tau^{\prime}}) \leq \frac{\beta \cdot \mathtt{LB}}{k} \leq \frac{\beta \cdot \mathtt{OPT}}{k} $.
  Therefore, we have
  \begin{align*}
    f(O \cup S_{\tau^{\prime}}) - f(S_{\tau^{\prime}})
    & \leq \sum_{o \in O \setminus S_{\tau^{\prime}}} \Delta_f(o|S_{\tau^{\prime}}) \\
    & = \sum_{o \in O_2} \Delta_f(o|S_{\tau^{\prime}}) + \sum_{o \in O_3} \Delta_f(o|S_{\tau^{\prime}}) \\
    & \leq \sum_{s \in S_{\tau^{\prime}}} \Delta_f(s|S^{\prime}) + \beta \cdot \mathtt{OPT} \\
    & = f(S_{\tau^{\prime}}) + \beta \cdot \mathtt{OPT}
  \end{align*}
  where $ S^{\prime} $ is the partial solution when $s$ is added to $ S_{\tau^{\prime}} $.
  And we conclude that $ f(S_{\tau^{\prime}}) \geq \frac{1-\beta}{2} \cdot \mathtt{OPT} $ from the above inequalities.
  In Case (2), we have $\tau^{\prime}$ is the maximum in $T$ and thus $ \tau^{\prime} \in [\frac{\delta_{max}}{1+\alpha}, \delta_{max}] $.
  We divide $O$ into $O_1,O_2,O_3$ in the same way as Case (1).
  It is easy to see that the results for $O_1$ and $O_3$ are exactly the same as Case (1).
  The only difference is that there may exist some items in $O_2$ rejected by $ S_{\tau^{\prime}} $ because their groups have been filled in $ S_{\tau^{\prime}} $.
  For any $ o \in O_2 $, we have $ \Delta_f(o|S_{\tau^{\prime}}) \leq \delta_{max} \leq (1+\alpha) \cdot \tau^{\prime} \leq (1+\alpha) \cdot \Delta_f(s|S^{\prime}) $ where $s$ is from the same group as $o$ and $S^{\prime}$ is the partial solution when $s$ is added.
  Accordingly, we can get $ \mathtt{OPT} - f(S_{\tau^{\prime}}) \leq (1+\alpha) \cdot f(S_{\tau^{\prime}})+ \beta \cdot \mathtt{OPT} $ and thus $ f(S_{\tau^{\prime}}) \geq \frac{1-\beta}{2+\alpha} \cdot \mathtt{OPT} $ in both cases.
\end{proof}

Next, we give the approximation ratio and complexity of SP-FSM in Theorems~\ref{thm:approx:op:fsm} and~\ref{thm:complexity:op:fsm}, respectively.

\begin{theorem}\label{thm:approx:op:fsm}
  Assuming that $ \alpha,\beta = O(\varepsilon) $, SP-FSM in Algorithm~\ref{alg:op:fsm} is a $ (\frac{1}{2}-\varepsilon) $-approximation algorithm for the FSM problem.
\end{theorem}
\begin{proof}
  According to the results of Lemmas~\ref{lm:approx:know:opt} and~\ref{lm:approx:post}, we have $ f(S) \geq \frac{1-\beta}{2+\alpha} \cdot \mathtt{OPT} $ for the solution $S$ returned by Algorithm~\ref{alg:op:fsm}.
  By assuming $ \alpha,\beta = O(\varepsilon) $, we conclude the proof.
\end{proof}

\begin{theorem}\label{thm:complexity:op:fsm}
  Assuming that $ \alpha,\beta = O(\varepsilon) $, \textnormal{SP-FSM} in Algorithm~\ref{alg:op:fsm} requires one pass over the data stream $V$, stores at most $ O\big( \frac{k \log{k}}{\varepsilon} + |B| \big) $ items, has $ O\big( \frac{\log{k}}{\varepsilon} \big) $ update time per item for stream processing, and takes $ O\big( \frac{k\log{k}}{\varepsilon} \cdot (|B|+k) \big) $ time for post-processing.
\end{theorem}
\begin{proof}
  The number $|T|$ of thresholds maintained at any time satisfies that $ (1+\alpha)^{|T|} \leq 2k $.
  Using the Taylor expansion of $\log(1+\alpha)$, we have $ |T| \leq \frac{\log{2k}}{\log{(1+\alpha)}} \leq \frac{\log{2k}}{\alpha\log{2}} = O\big( \frac{\log{k}}{\alpha} \big) $.
  Therefore, the number of function evaluations per item is $ O\big( \frac{\log{k}}{\varepsilon} \big) $.
  Since each candidate solution contains at most $k$ items, the total number of items stored in SP-FSM is $ O\big( \frac{k \log{k}}{\varepsilon} + |B| \big) $.
  For each candidate solution $ S_{\tau} $, the post-processing procedure runs in $(k-|S_{\tau}|)$ iterations and processes at most $ (|B| + k) $ items at each iteration.
  Therefore, it takes $ O\big( \frac{k\log{k}}{\varepsilon} \cdot (|B|+k) \big) $ time for post-processing.
\end{proof}

\begin{figure*}[ht]
  \centering
  \includegraphics[width=0.75\textwidth]{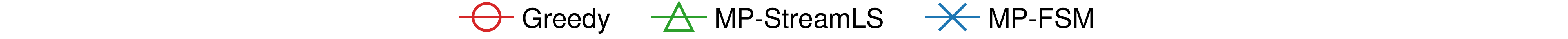}
  \\
  \subfigure[POKEC (Gender, PR)]{
    \includegraphics[width=0.23\textwidth]{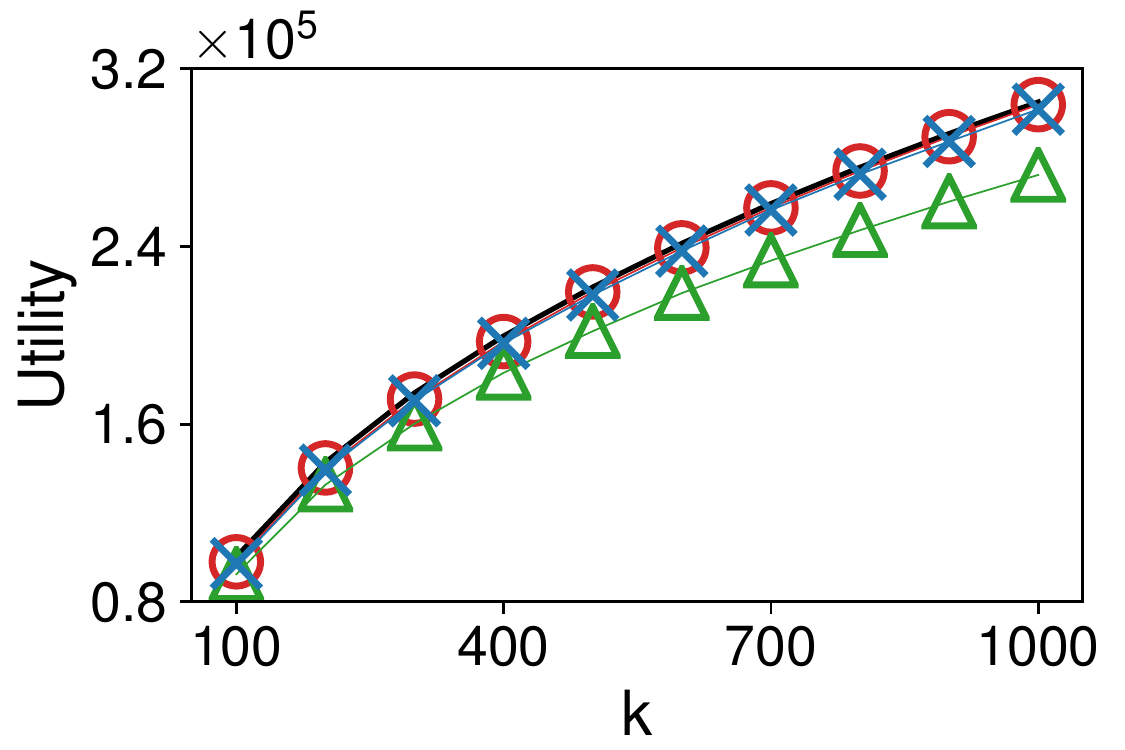}
  }
  \subfigure[POKEC (Gender, ER)]{
    \includegraphics[width=0.23\textwidth]{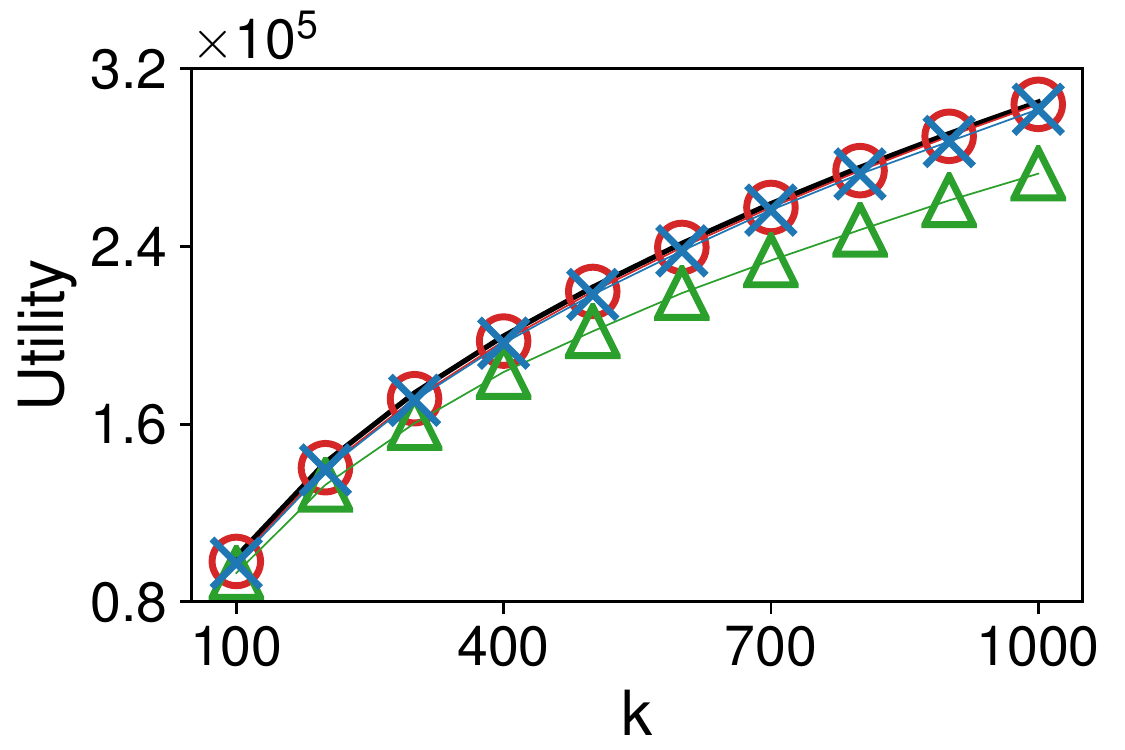}
  }
  \subfigure[POKEC (Age, PR)]{
    \includegraphics[width=0.23\textwidth]{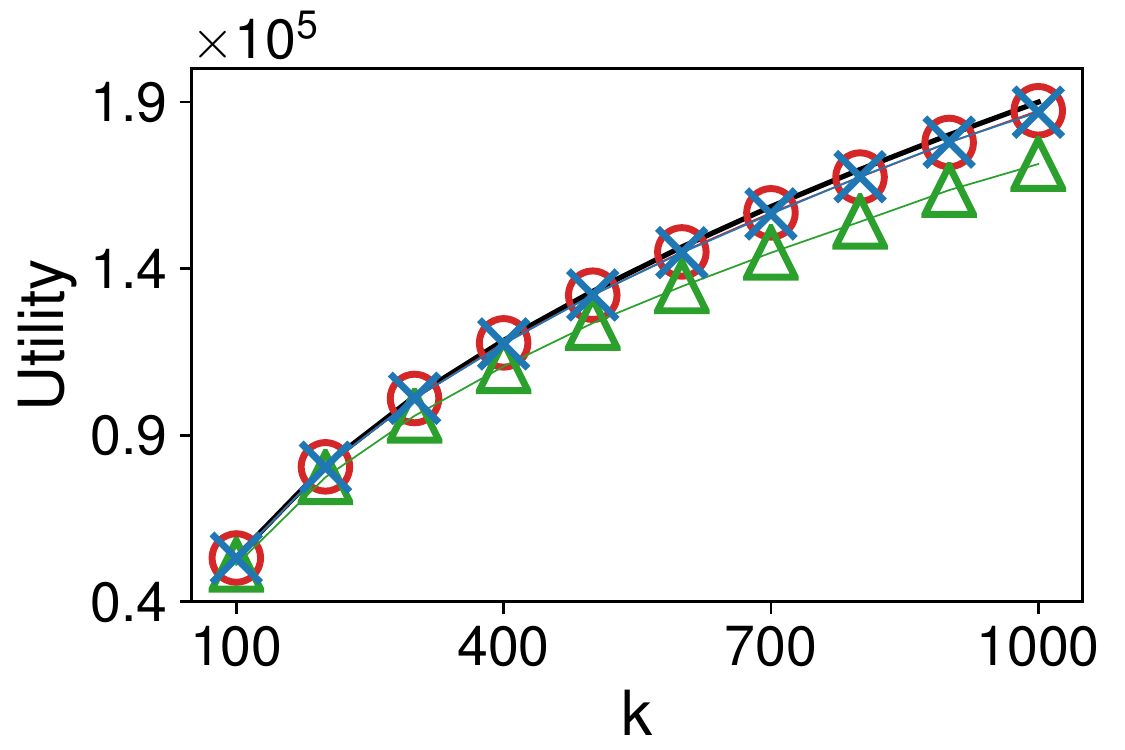}
  }
  \subfigure[POKEC (Age, ER)]{
    \includegraphics[width=0.23\textwidth]{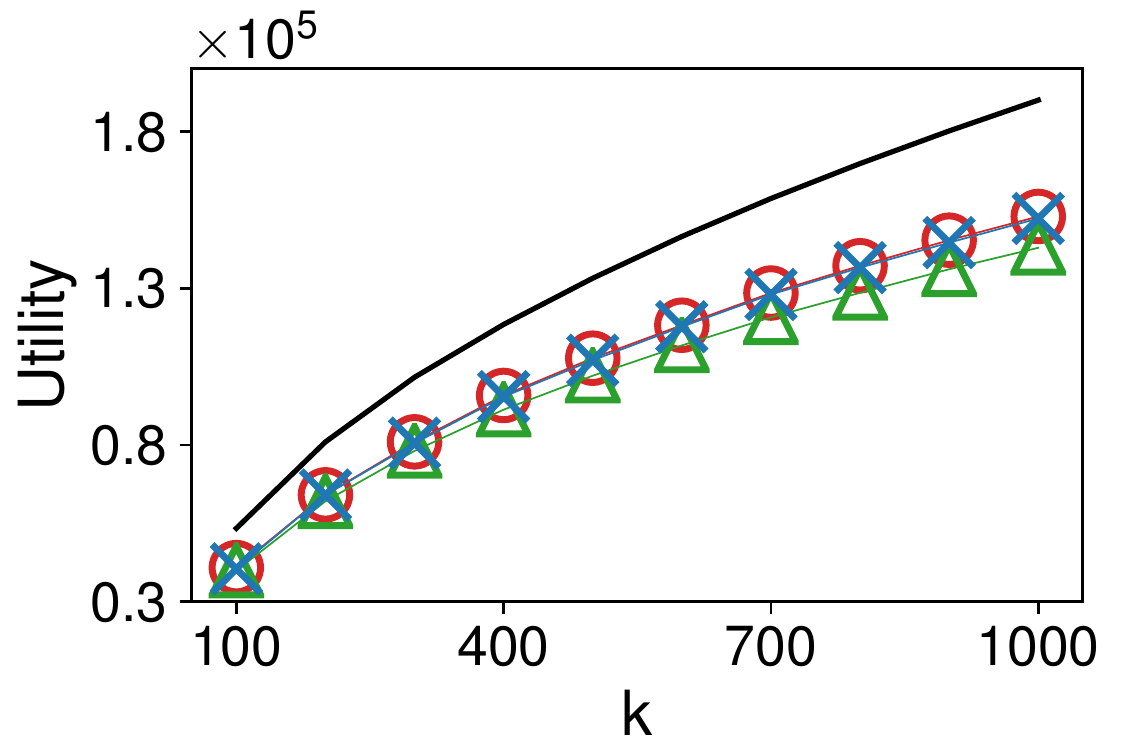}
    \label{subfig:price:fairness}
  }
  \caption{Solution utilities of multi-pass algorithms on POKEC. The results of \textsc{Greedy} without any fairness constraint are plotted as black lines to illustrate ``the prices of fairness''.}
  \label{fig:mc-mp-k-util}
  \Description{utility, max cover, multi-pass, k}
\end{figure*}
\begin{figure*}[ht]
  \centering
  \includegraphics[width=0.75\textwidth]{figs/legend-mp.pdf}
  \\
  \subfigure[POKEC (Gender, PR)]{
    \includegraphics[width=0.23\textwidth]{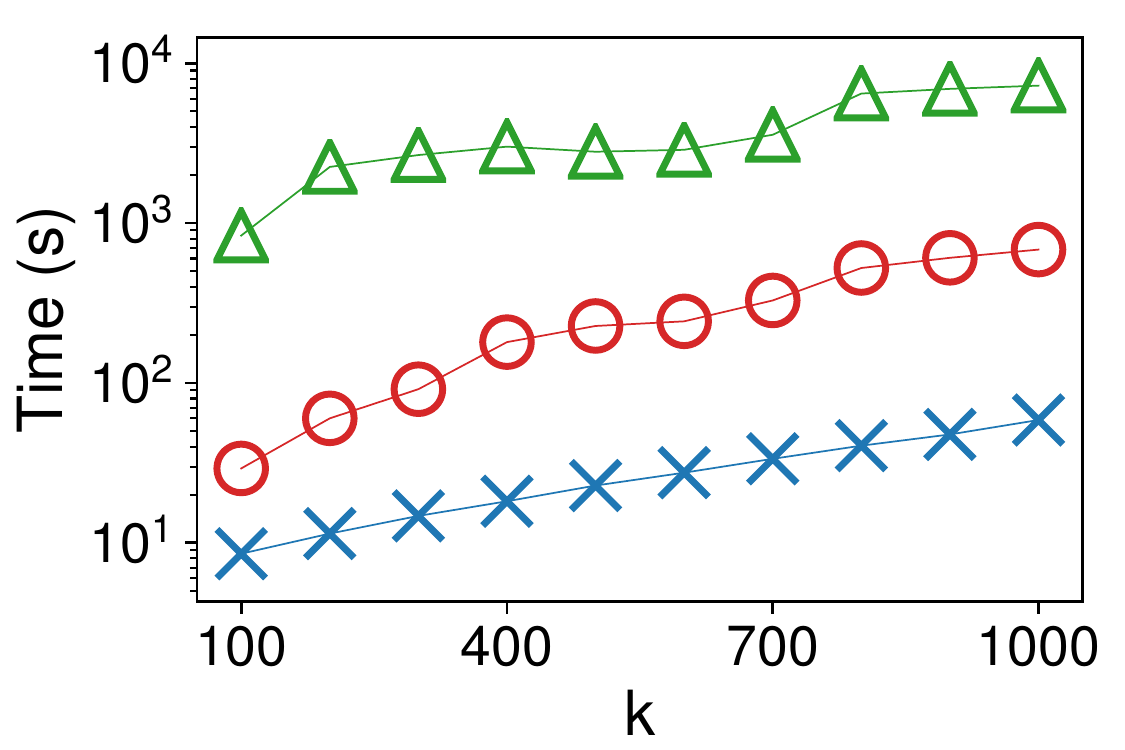}
  }
  \subfigure[POKEC (Gender, ER)]{
    \includegraphics[width=0.23\textwidth]{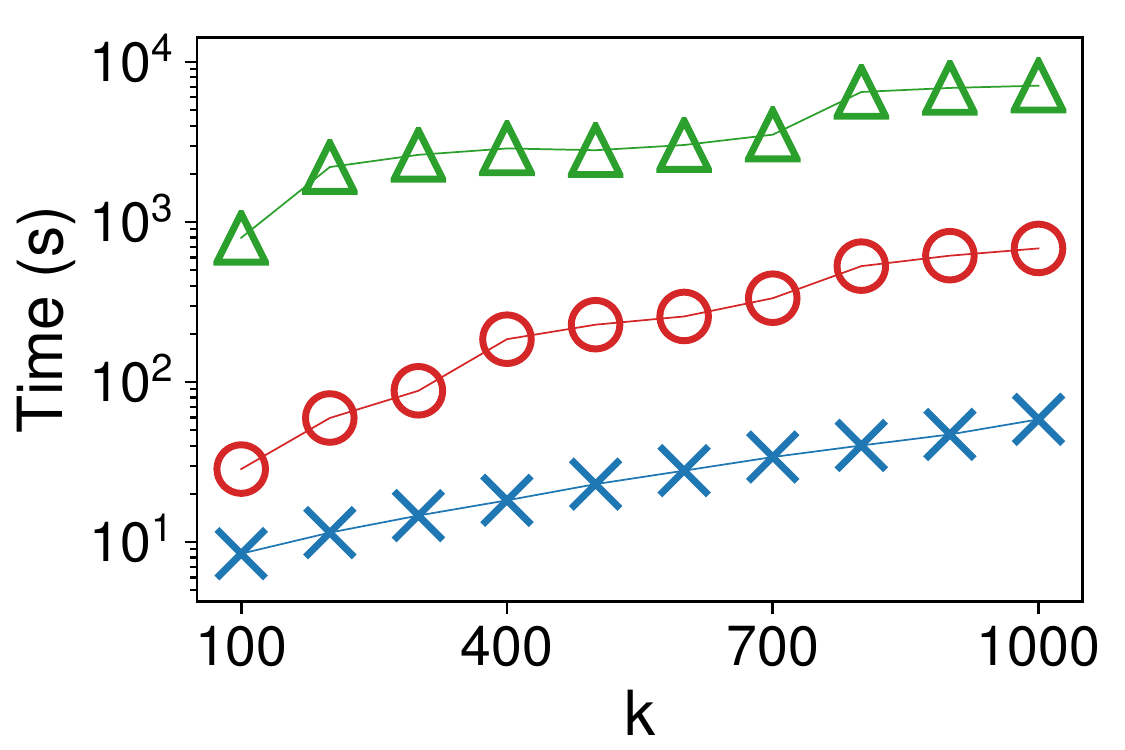}
  }
  \subfigure[POKEC (Age, PR)]{
    \includegraphics[width=0.23\textwidth]{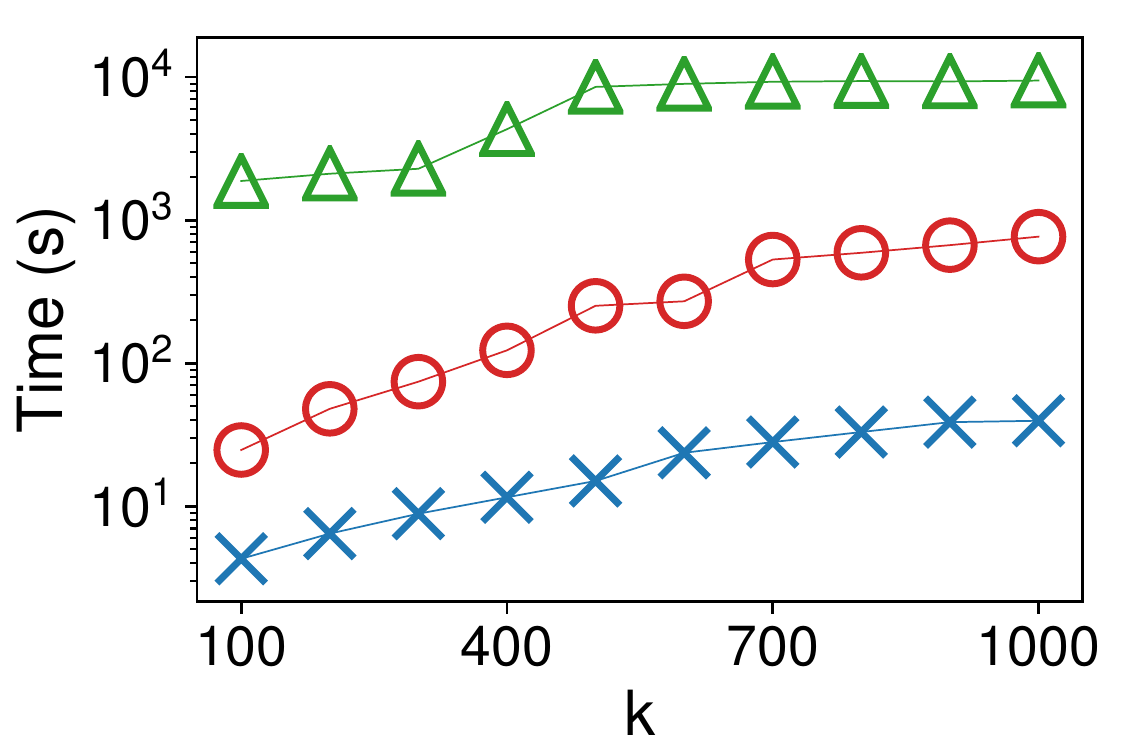}
  }
  \subfigure[POKEC (Age, ER)]{
    \includegraphics[width=0.23\textwidth]{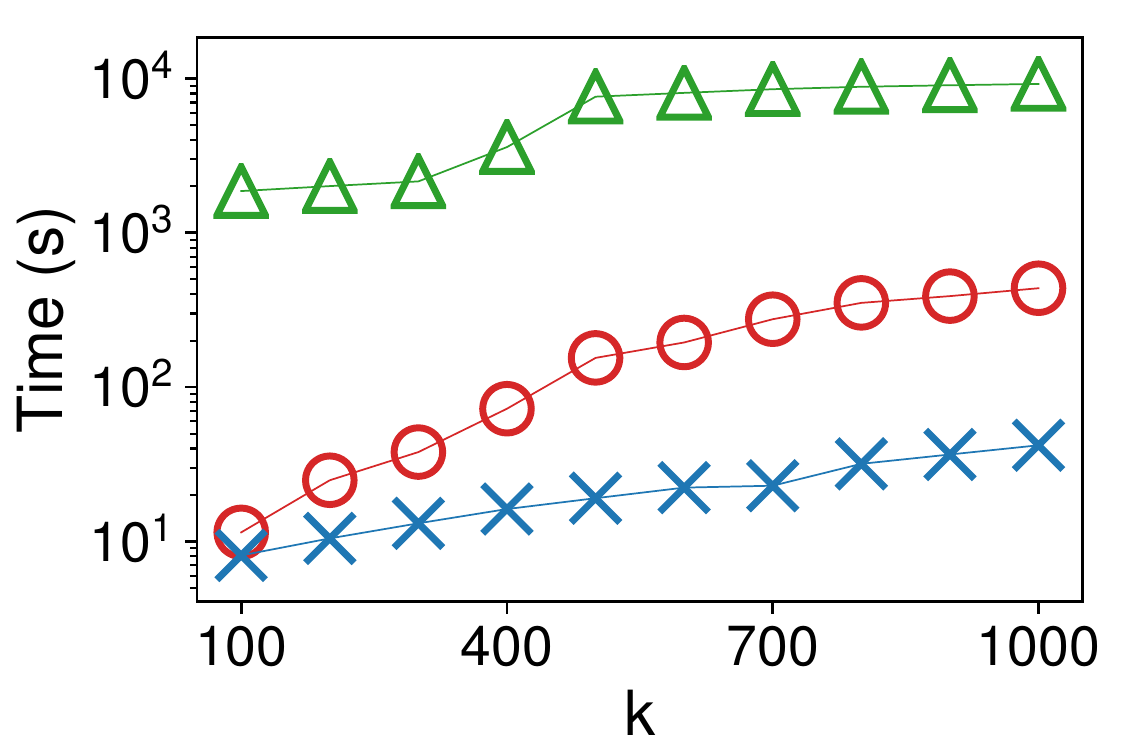}
  }
  \caption{Running time of multi-pass algorithms on POKEC.}
  \label{fig:mc-mp-k-time}
  \Description{time, max cover, multi-pass, k}
\end{figure*}

\subsection{SP-FSM with Bounded Buffer Size}
\label{sec:small_buffer}

From the above results, we can see that SP-FSM may store $O(n)$ items in the buffer and take $O\big( \frac{nk\log{k}}{\varepsilon} \big)$ time for post-processing in the worst case.
In practice, a streaming algorithm is often required to process massive data streams with limited time and memory (sublinear to or independent of $n$).
And it is not favorable for SP-FSM to store an unlimited number of items in the buffer $B$.
Therefore, we propose a simple strategy for SP-FSM to manage the buffered items so that the buffer size is always bounded at the expense of lower approximation ratios in adversary settings.

We consider that the maximum buffer size is restricted to $k^{\prime} = O(k)$ and extra items should be dropped from $B$ once its size exceeds $k^{\prime}$.
The following rules are considered for buffer management.
Firstly, since $\mathtt{LB}$ increases over time, it is safe to drop at any time during stream processing any item already in the buffer whose marginal gain is lower than $\frac{\beta \cdot \mathtt{LB}}{k}$ for the current value of $\mathtt{LB}$, without affecting the theoretical guarantee.
Secondly, to avoid duplications, if an item is added to some candidate solution but needs to be buffered for another, it is not necessary to add this item to the buffer because the algorithm has already stored this item.
In this case, items in both candidates and the buffer should be used for post-processing.
Thirdly, as the buffer is used for storing high-utility items for post-processing, the items with larger marginal gains should have higher priorities to be stored. 
If the buffer size still exceeds $k^{\prime}$ after (safely) dropping items using the first two rules, it is required to sort the items in $B$ in a descending order of marginal gain $ \delta(v) = \max_{\tau \in T} \Delta_f(v,S_{\tau}) $ and drop the item $v$ with the lowest $\delta(v)$ until $ |B|=k^{\prime} $. 
Fourthly, considering the fairness constraint, it will not drop any item $v$ from $ V_i $ anymore if $ |B \cap V_i| \leq k_i $ even if $\delta(v)$ is among the lowest marginal gains.
In this case, it will drop the item with the lowest $\delta(v)$ from $ V_i $ with $ |B \cap V_i| > k_i $ instead.

The first two rules above have no effect on the theoretical guarantee on the approximation ratio of SP-FSM.
The latter two rules will lower the approximation ratio of SP-FSM in some cases.
Let $v^{\prime}$ be the item with the largest $\delta(v)$ among all items dropped due to Rule (3) or (4).
The approximation ratio of SP-FSM will drop to $ \frac{1-\beta^{\prime}}{2} $ where $ \beta^{\prime} = \frac{k \cdot \delta(v^{\prime})}{\mathtt{LB}} $.
Once $ \beta^{\prime} \geq 1 - \frac{1}{k} $, the approximation ratio will become $\frac{1}{2k}$ in the worst case.
Nevertheless, according to our experimental results in Section~\ref{sec:exp}, SP-FSM provides high-quality solutions empirically with very small buffer sizes (i.e., $k^{\prime}=2k$).

%!TEX root = main.tex
\section{Experiments}\label{sec:exp}

\begin{figure*}[ht]
  \includegraphics[width=0.75\textwidth]{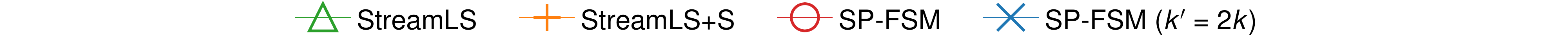}
  \\
  \subfigure[POKEC (Gender, PR)]{
    \includegraphics[width=0.23\textwidth]{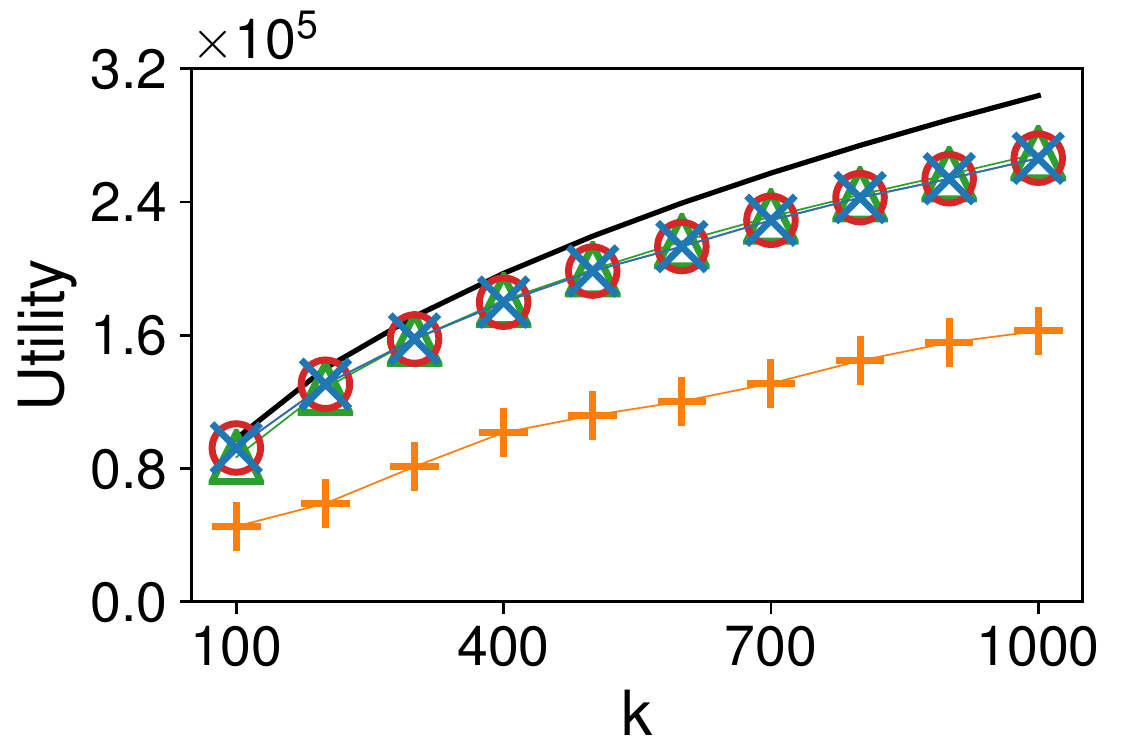}
  }
  \subfigure[POKEC (Gender, ER)]{
    \includegraphics[width=0.23\textwidth]{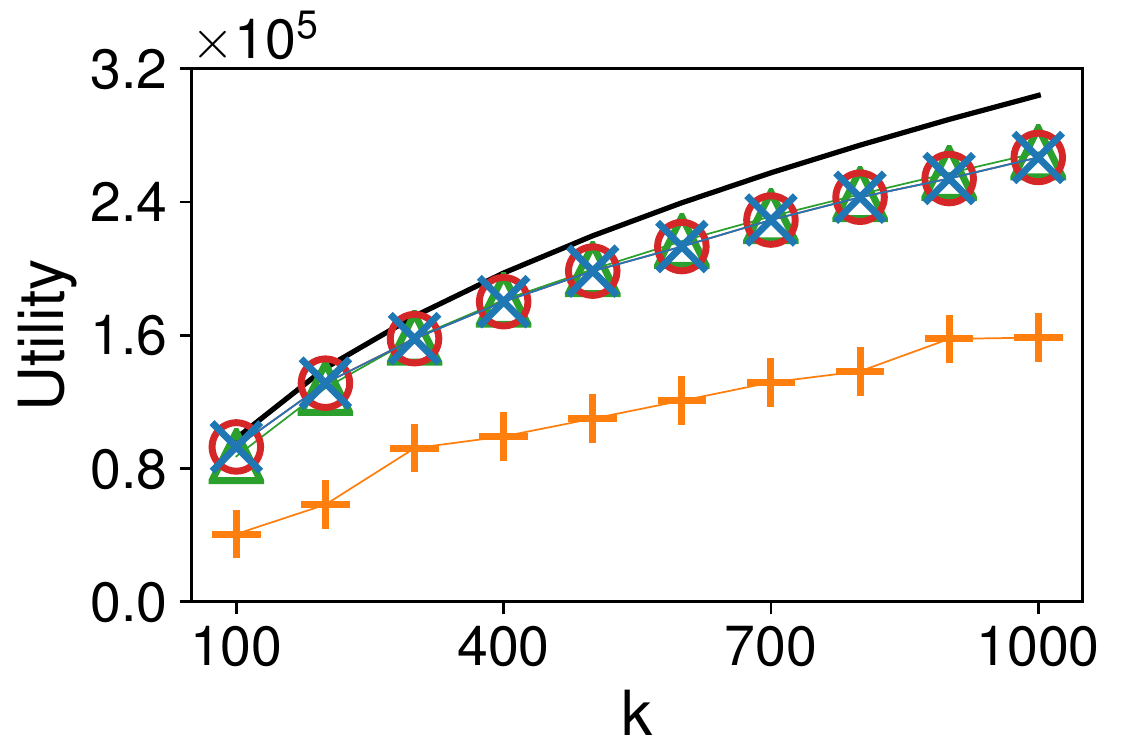}
  }
  \subfigure[POKEC (Age, PR)]{
    \includegraphics[width=0.23\textwidth]{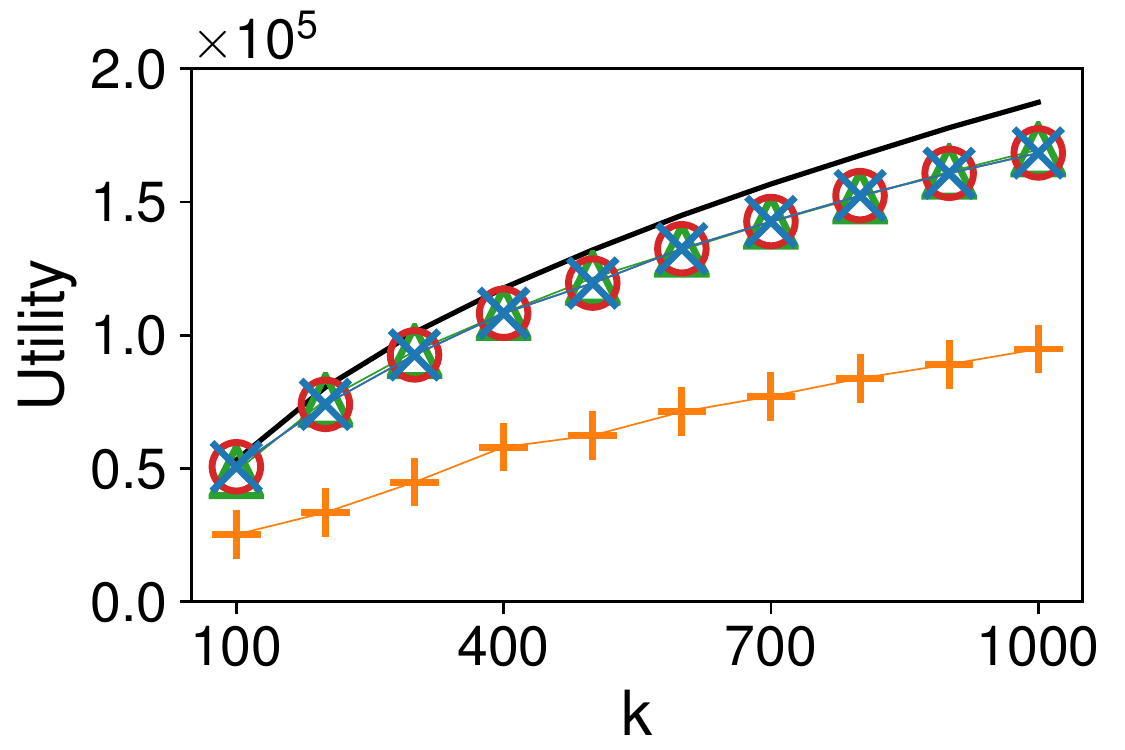}
  }
  \subfigure[POKEC (Age, ER)]{
    \includegraphics[width=0.23\textwidth]{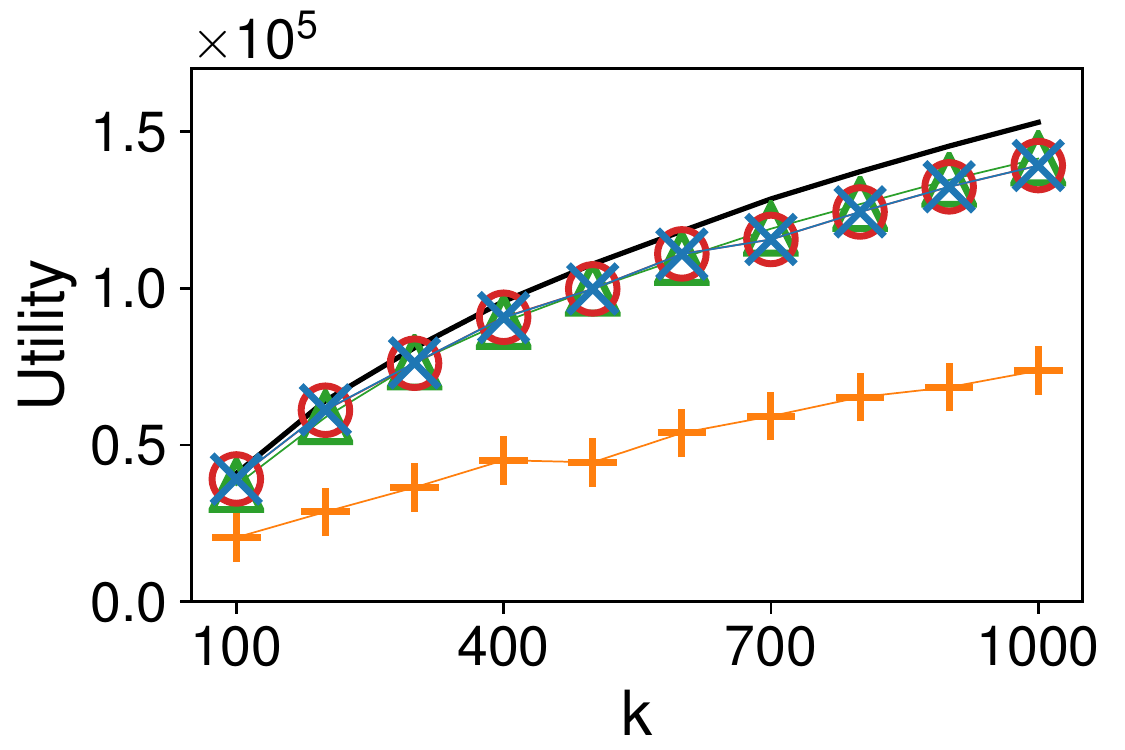}
  }
  \caption{Solution utilities of single-pass algorithms on POKEC. The results of \textsc{Greedy} are plotted as black lines to illustrate ``the prices of streaming data access''.}
  \label{fig:mc-sp-k-util}
  \Description{utility, max cover, single-pass, k}
\end{figure*}
\begin{figure*}[ht]
  \includegraphics[width=0.75\textwidth]{figs/legend-sp.pdf}
  \\
  \subfigure[POKEC (Gender, PR)]{
    \includegraphics[width=0.23\textwidth]{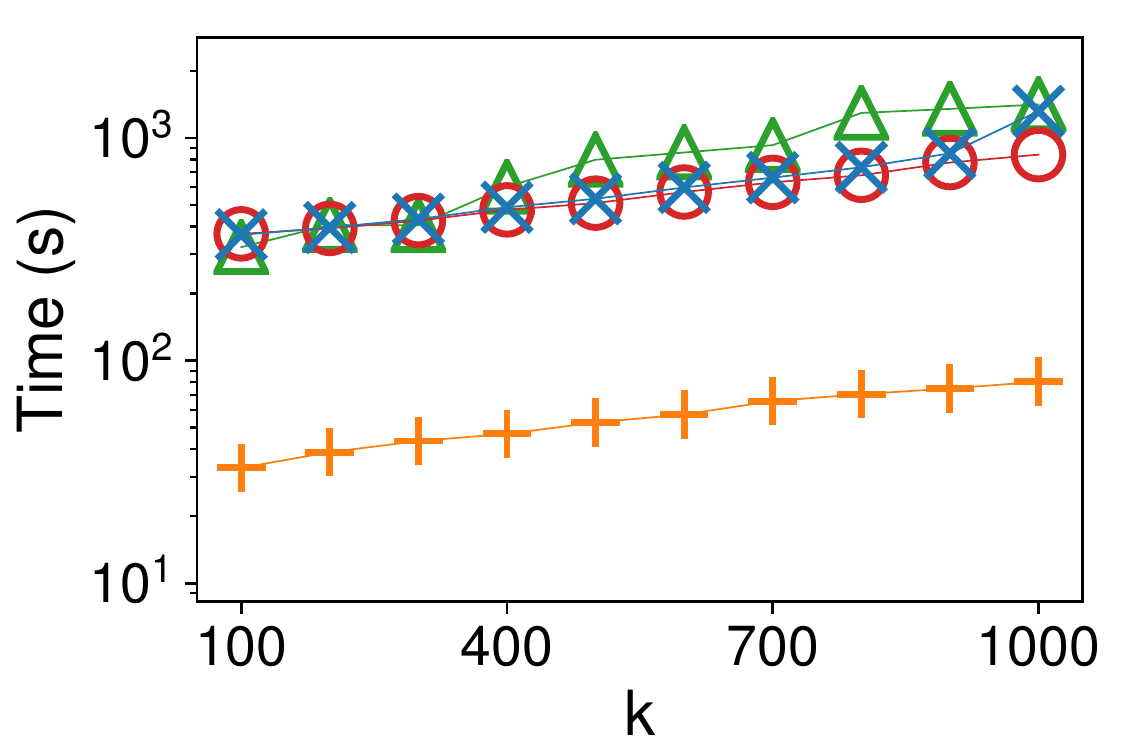}
  }
  \subfigure[POKEC (Gender, ER)]{
    \includegraphics[width=0.23\textwidth]{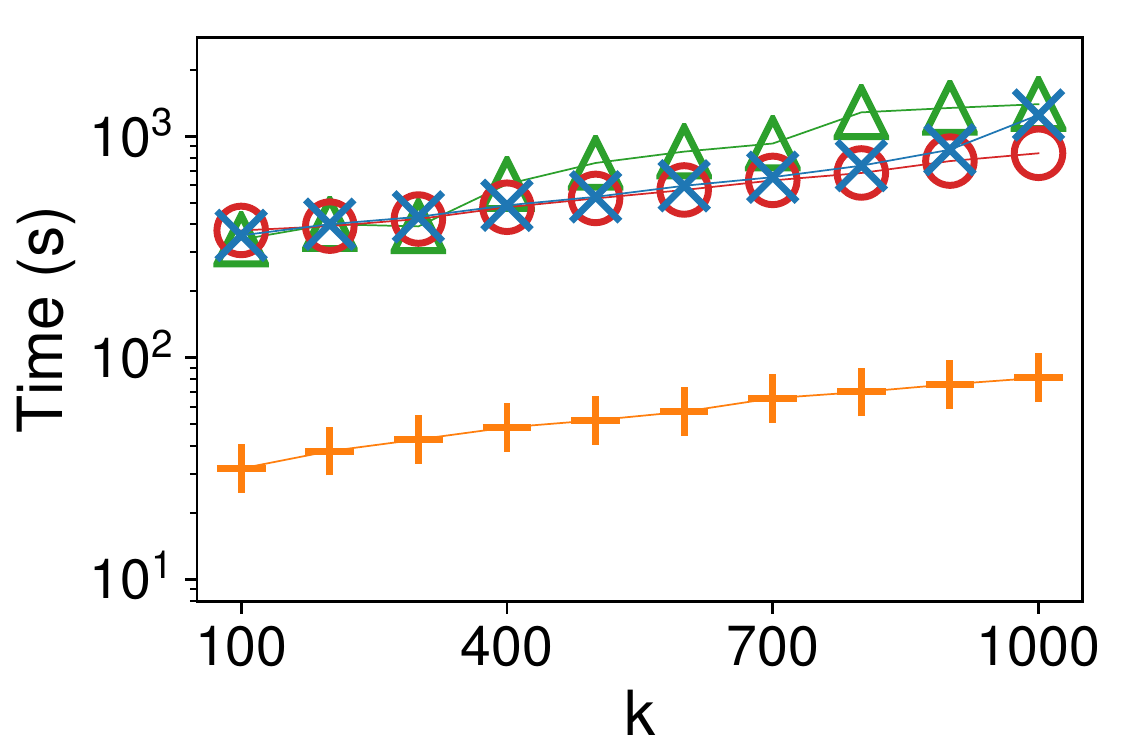}
  }
  \subfigure[POKEC (Age, PR)]{
    \includegraphics[width=0.23\textwidth]{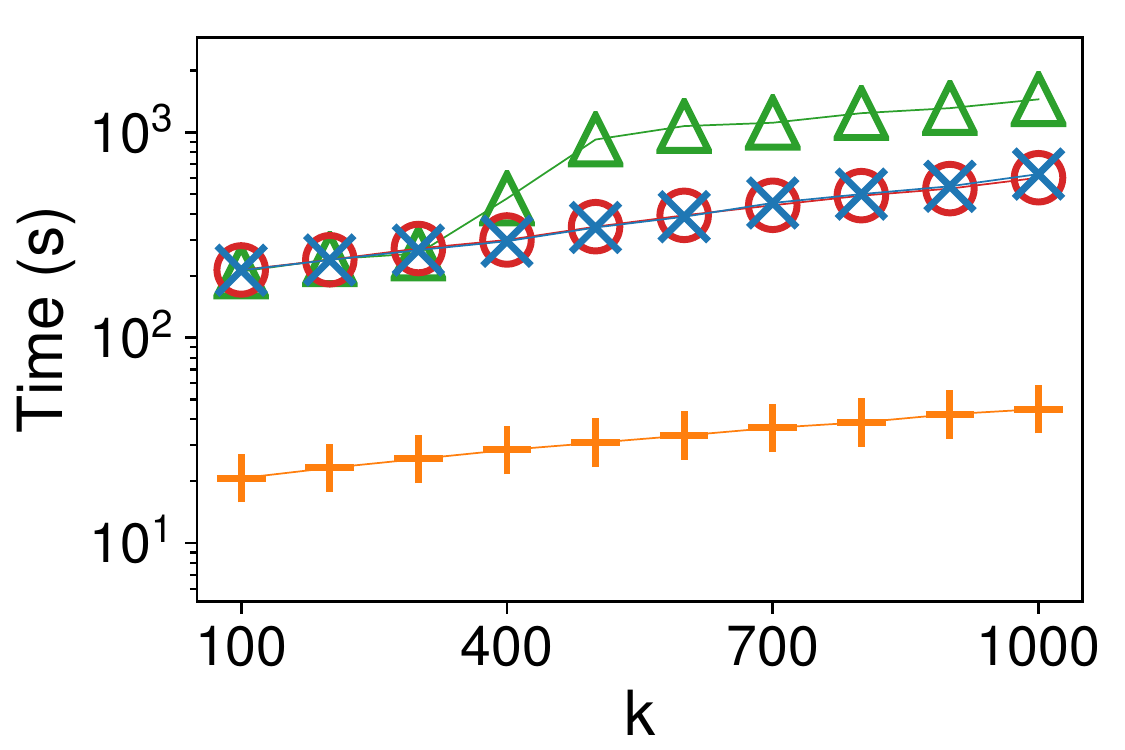}
  }
  \subfigure[POKEC (Age, ER)]{
    \includegraphics[width=0.23\textwidth]{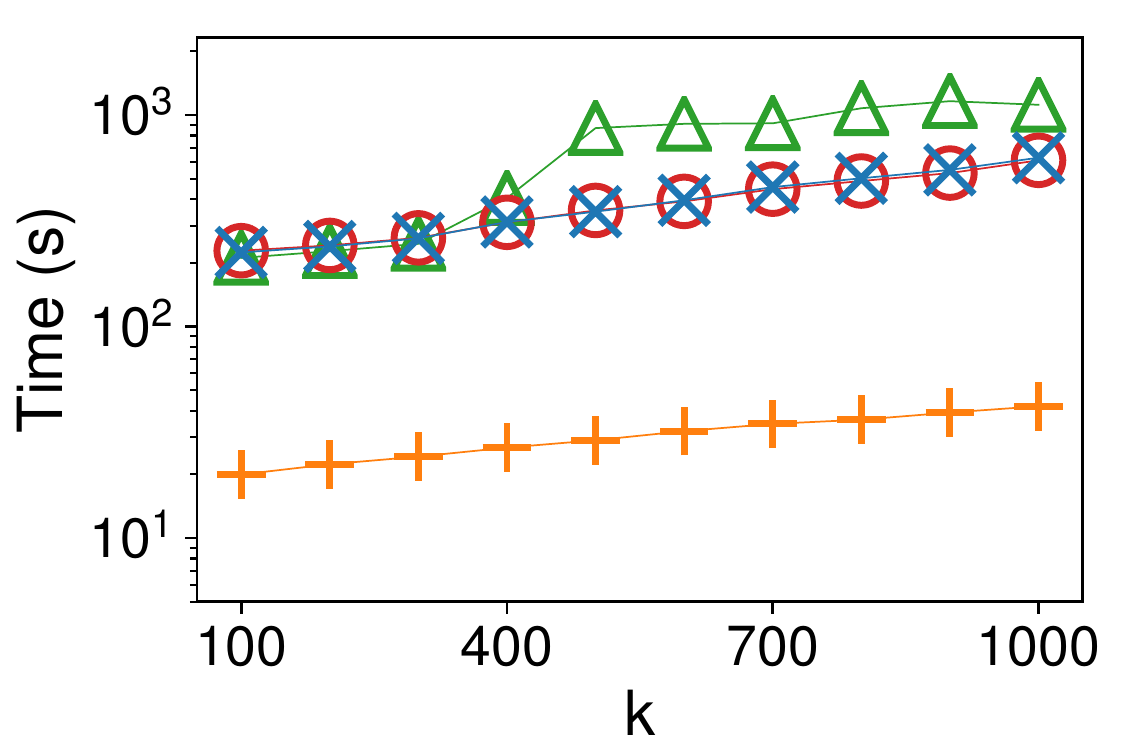}
  }
  \caption{Running time of single-pass algorithms on POKEC.}
  \label{fig:mc-sp-k-time}
  \Description{time, max cover, single-pass, k}
\end{figure*}

The goal of our experiments is three-fold.
First, we aim to quantify ``the prices of fairness and streaming'', i.e., the losses in solution utilities
caused by introducing the fairness constraint and restricting data access to a single pass over the stream.
Second, we aim to demonstrate the improvements of MP-FSM upon existing algorithms in the multi-pass streaming setting.
Third, we aim to illustrate that SP-FSM (with unlimited and bounded buffer sizes) outperforms
existing single-pass streaming algorithms.

Towards this end, we perform extensive experiments on two applications, namely \emph{maximum coverage on large graphs}
and \emph{personalized recommendation}, for evaluation.
We compare MP-FSM with the following two multi-pass streaming algorithms:
\begin{itemize}
  \item \textsc{Greedy}:
  the classic $\frac{1}{2}$-approximation $k$-pass greedy algorithm
  proposed by Fisher et al.~\cite{Fisher1978}.
  \item \textsc{MP-StreamLS}:
  a $\frac{1}{2+\varepsilon}$-approximation $ O(\frac{1}{\varepsilon}) $-pass streaming algorithm
  in~\cite{DBLP:conf/approx/HuangTW20}.
\end{itemize}
Moreover, we compare SP-FSM with the following two single-pass streaming algorithms:
\begin{itemize}
  \item \textsc{StreamLS}:
  a $\frac{1}{4}$-approximation streaming algorithm in~\cite{DBLP:journals/mp/ChakrabartiK15,DBLP:conf/icalp/ChekuriGQ15}.
  \item \textsc{StreamLS+S}:
  an improved version of \textsc{StreamLS} with subsampling in~\cite{DBLP:conf/nips/FeldmanK018}.
  The subsampling rate $q$ is set to $0.1$.
\end{itemize}

All algorithms were implemented in Python 3.6, and the experiments were conducted on a server running Ubuntu 16.04 with an Intel Broadwell 2.40GHz CPU and 29GB memory.
Our implementation is publicly available on GitHub\footnote{\url{https://github.com/FraFabbri/fair-subset-datastream}}.
For each of the experiments, we invoked our algorithms with the following parameter values:
MP-FSM with $ \varepsilon=0.2 $ and SP-FSM with $ \alpha,\beta=0.5 $ and $k^{\prime}=2k$
in all cases where the buffer size is bounded. Note that in all the following figures,
we refer to SP-FSM with an unlimited buffer size as \textbf{SP-FSM} and SP-FSM with $k^{\prime}=2k$
as \textbf{SP-FSM ($k^{\prime}=2k$)}.

\subsection{Maximum Coverage on Large Graphs}\label{subsec:exp:mkc}

Maximum coverage is a classic submodular optimization task on graphs with many real-world applications such as community detection~\cite{DBLP:journals/datamine/GalbrunGT14}, influence maximization~\cite{DBLP:journals/pvldb/WangFLT17}, and web monitoring~\cite{DBLP:conf/sdm/SahaG09}.
The goal of this task is to select a small subset of nodes that covers a large portion of nodes in a graph.
Formally, given a graph $G=(V,E)$ where $n=|V|$ is the number of nodes and $m=|E|$ is the number of edges,
the goal is to find a size-$k$ subset $ S $ of $ V $ that maximizes the nodes in the neighborhood of $S$,
i.e., $ f(S) = |\bigcup_{v \in S} N(v)| $ where $N(v)$ is the set of nodes connected to $v$.
It is easy to verify that $f$ is nonnegative, monotone, and submodular.

We perform the experiments for maximum coverage on two graph datasets as follows:
(1) \textbf{POKEC} is a real-world dataset published on SNAP\footnote{\url{https://snap.stanford.edu/data/soc-Pokec.html}}.
It is a directed graph with 1,632,803 nodes and 30,622,564 edges representing the follower/followee relationships among users in Pokec.
Each node is associated with a user profile with demographic information.
The nodes are partitioned into $l=2$ groups by gender or $l=7$ groups by age in our experiments.
(2) \textbf{SYN} is a set of synthetic graphs generated by the Barab{\'{a}}si-Albert model~\cite{RevModPhys.74.47} with equal number of nodes and edges, i.e.,~$n=m$.
To test the effect of graph size, we generate different graphs by ranging $n$ from $100$k to $1$m.
The nodes are randomly partitioned into $l$ groups and the group sizes follow a Zipf's distribution with parameter $s=2$.
By default, we set the number $l$ of groups to $10$.
To test the effect of $l$, we fix $n=500$k and vary $l$ from $10$ to $100$.

In the first set of experiments, we evaluate the performance of \textsc{Greedy}, \textsc{MP-StreamLS},
and MP-FSM in the multi-pass streaming setting. We range the total cardinality constraint $k = \sum_{i} k_i$
from $100$ to $1,000$ and use both \emph{proportional representation} (PR) and \emph{equal representation} (ER) to assign the group-specific cardinality constraint $k_i$ for each $ i \in [l] $.
The solution utilities and running time on POKEC are presented in Figures~\ref{fig:mc-mp-k-util}
and~\ref{fig:mc-mp-k-time}, respectively.
``The price of fairness'' -- i.e., the loss in utility caused by the fairness constraint,
is marginal for PR in both cases of gender and age groups,
and ER in the case of gender groups, as two gender groups are roughly balanced
(e.g., $51\%$ female vs.~$49\%$ male) on POKEC.
However, for highly imbalanced groups (e.g., age groups on POKEC), enforcing equal representation
leads to significant losses in utilities (see Figure~\ref{subfig:price:fairness}).
MP-FSM outperforms \textsc{Greedy} and \textsc{MP-StreamLS} in terms of both running time
and solution utility in almost all cases.
It runs up to $19$ and $567$ times faster than \textsc{Greedy} and \textsc{MP-StreamLS}, respectively.
Meanwhile, its solution utilities are always nearly equal to (at least $99\%$ of) those of \textsc{Greedy}
and consistently (up to $10\%$) higher than those of \textsc{MP-StreamLS}.

\begin{figure*}[ht]
  \centering
  \includegraphics[width=0.75\textwidth]{figs/legend-mp.pdf}
  \\
  \subfigure[Varying $l$]{
    \includegraphics[width=0.23\textwidth]{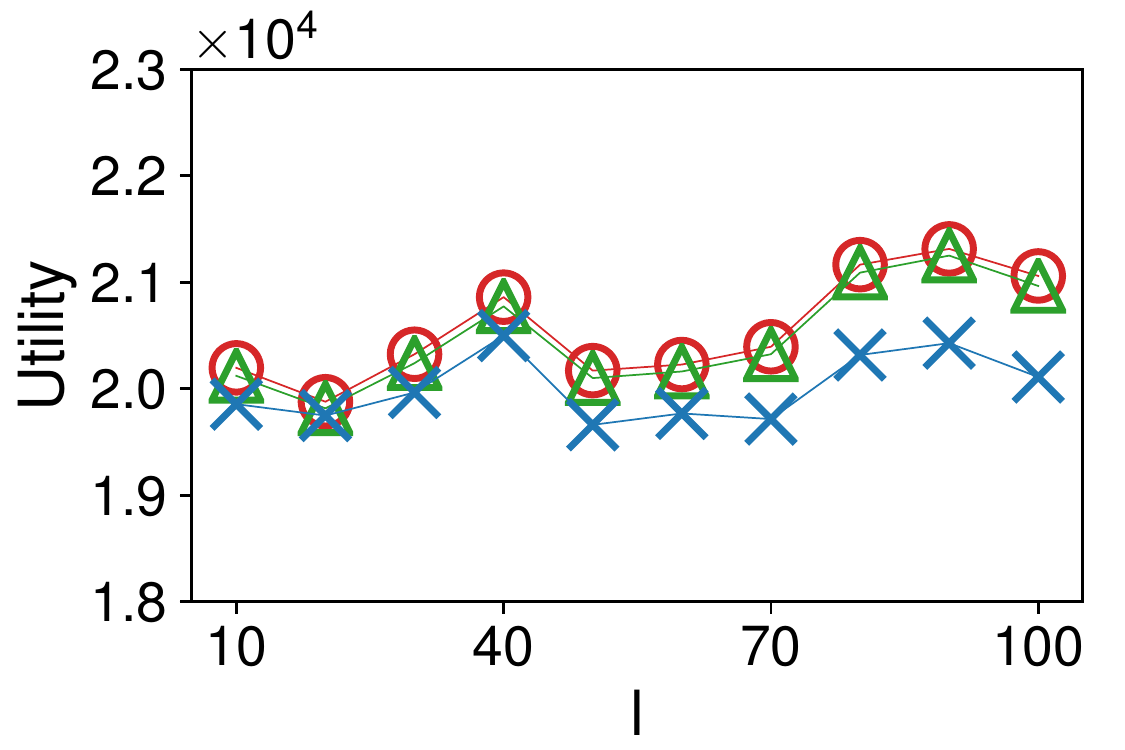}
    \includegraphics[width=0.23\textwidth]{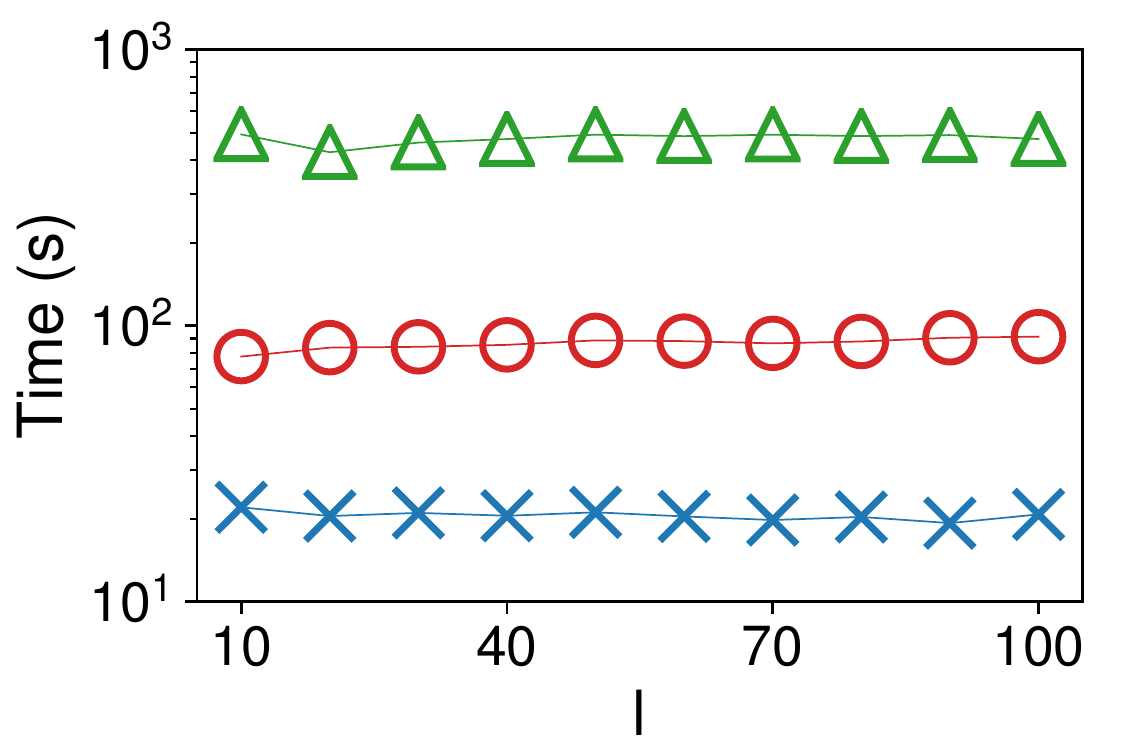}
  }
  \subfigure[Varying $n$]{
    \includegraphics[width=0.23\textwidth]{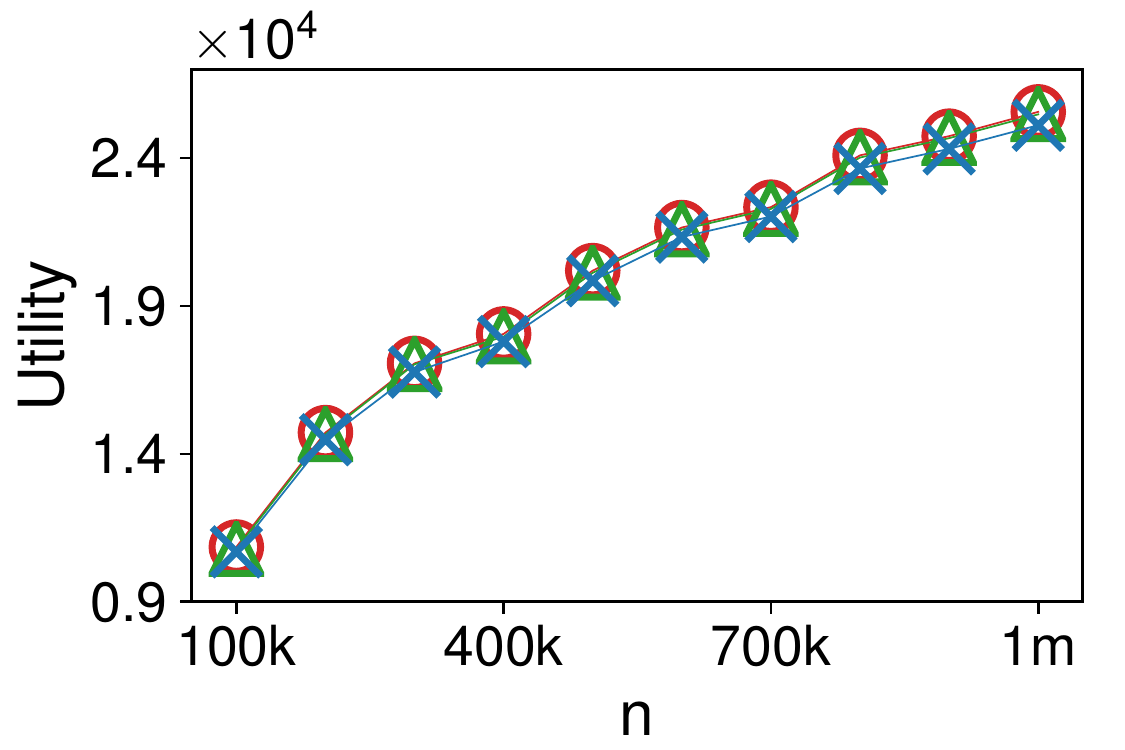}
    \includegraphics[width=0.23\textwidth]{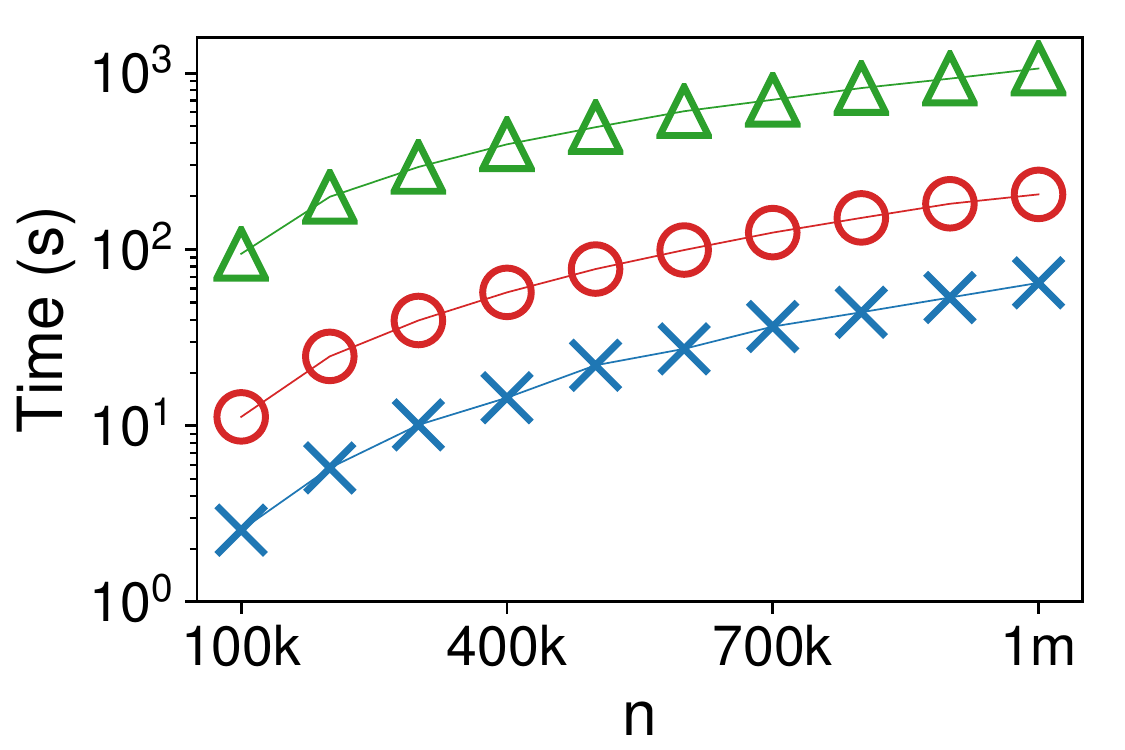}
  }
  \caption{Performance of multi-pass algorithms on SYN with varying dataset size $n$ and number of groups $l$.}
  \label{fig:mc-mp-nl}
  \Description{multi-pass, max cover, n, l}
\end{figure*}
\begin{figure*}[ht]
  \centering
  \includegraphics[width=0.75\textwidth]{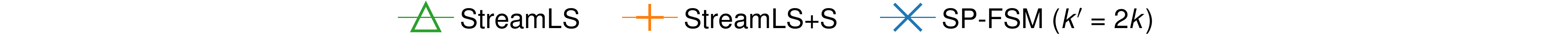}
  \\
  \subfigure[Varying $l$]{
    \includegraphics[width=0.23\textwidth]{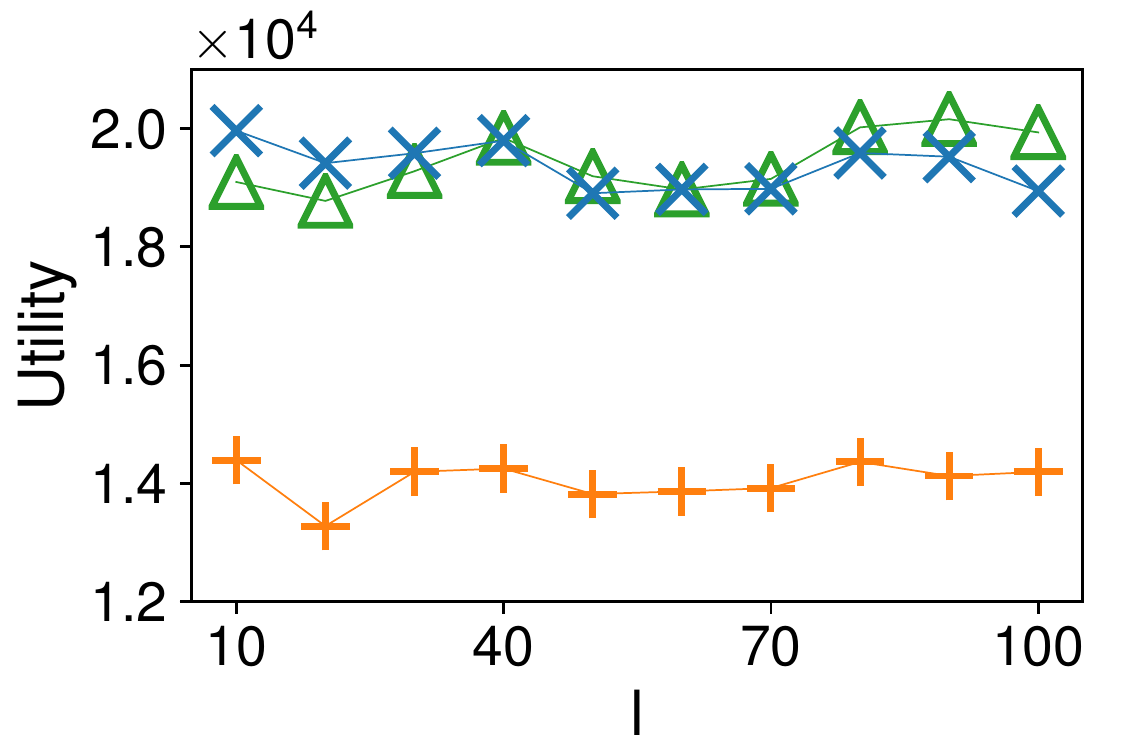}
    \includegraphics[width=0.23\textwidth]{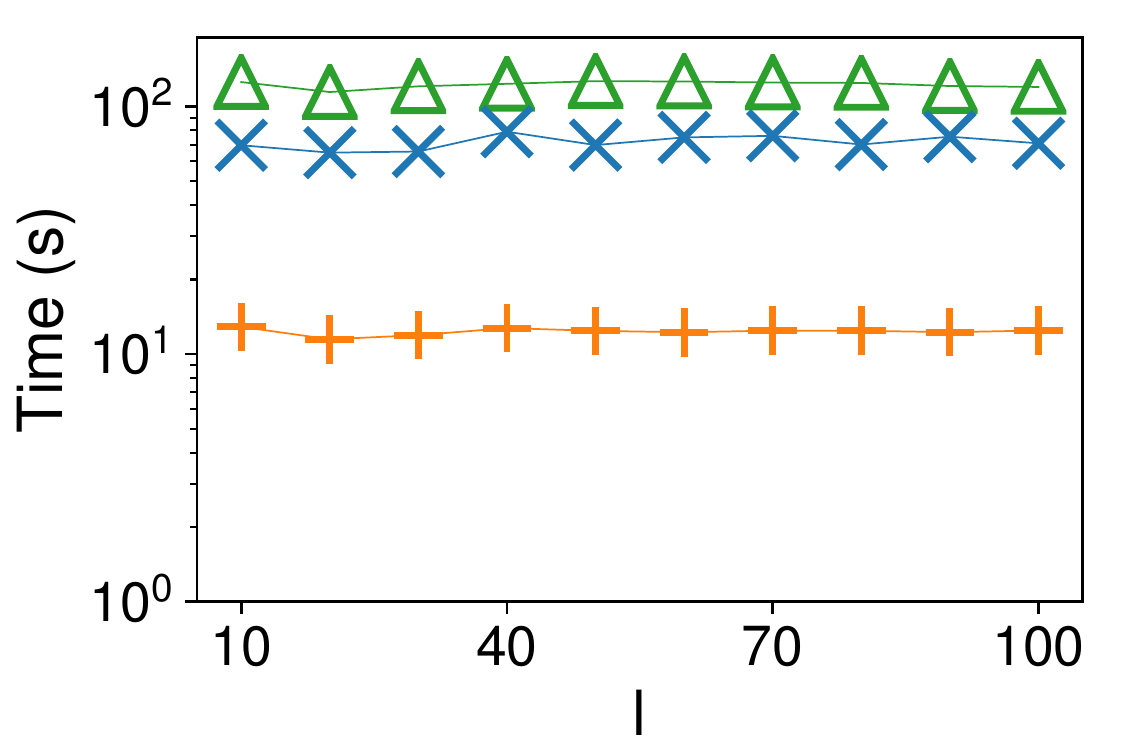}
  }
  \subfigure[Varying $n$]{
    \includegraphics[width=0.23\textwidth]{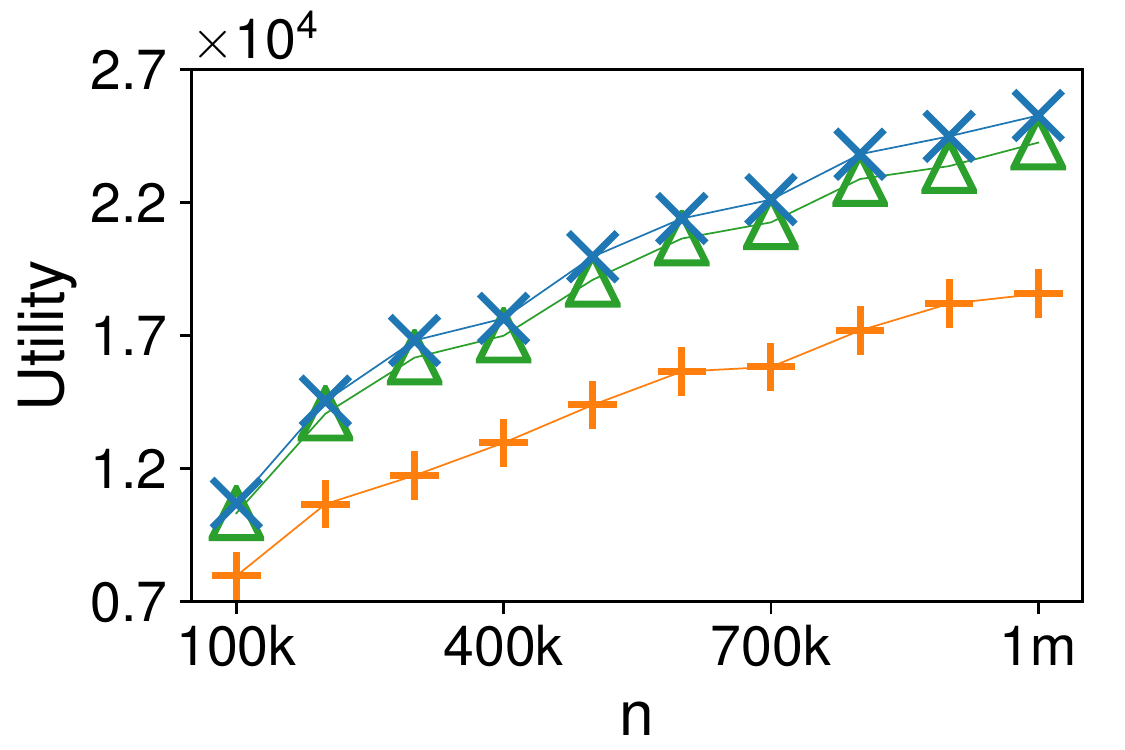}
    \includegraphics[width=0.23\textwidth]{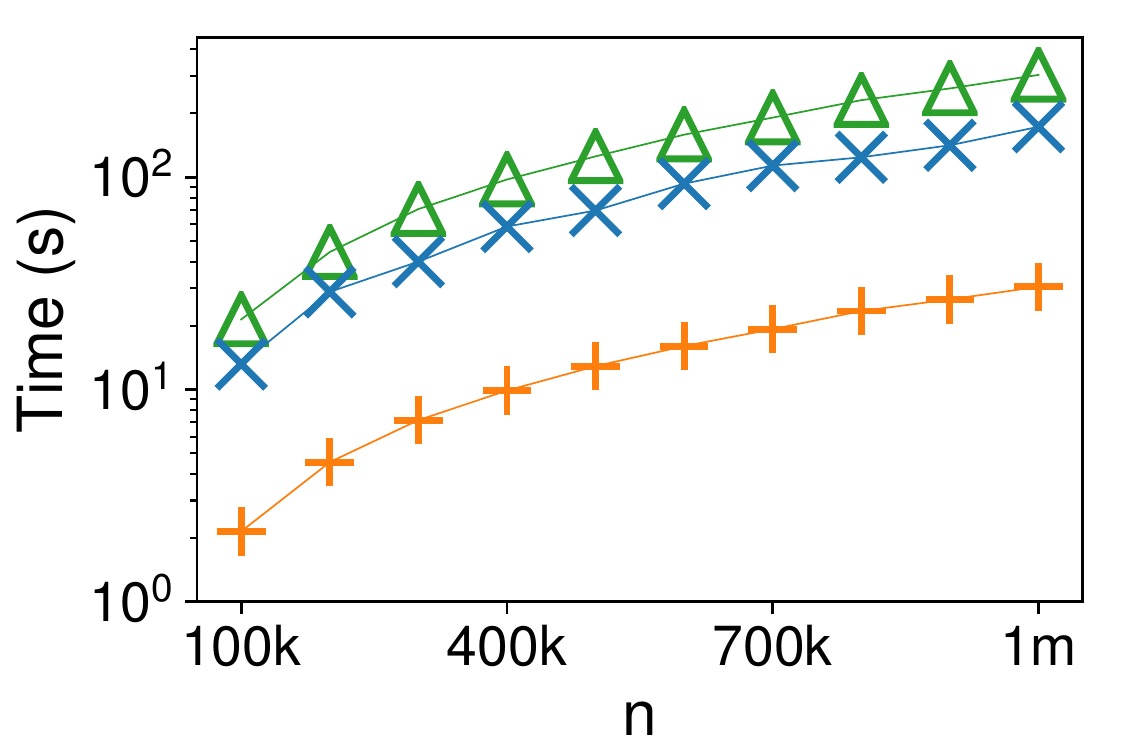}
  }
  \caption{Performance of single-pass algorithms on SYN with varying dataset size $n$ and number of groups $l$.}
  \label{fig:mc-sp-nl}
  \Description{single-pass, max cover, n, l}
\end{figure*}

In the second set of experiments, we evaluate the performance of \textsc{StreamLS}, \textsc{StreamLS+S},
and SP-FSM with unlimited and bounded (i.e., $ k^\prime=2k $) buffer sizes in the single-pass streaming setting.
We also vary $k$ from $100$ to $1,000$ and use both PR and ER for fairness constraints.
The experimental results on POKEC are illustrated in Figures~\ref{fig:mc-sp-k-util} and~\ref{fig:mc-sp-k-time}.
Firstly, the utilities of the solutions provided by \textsc{StreamLS} and SP-FSM are
typically around $10\%$ lower than the utilities of the solutions of \textsc{Greedy}.
This can be seen as ``the price of streaming data access'' -- i.e., the loss in utility for restricting data access only to a single pass over the stream.
Secondly, the solution quality of SP-FSM is generally equivalent to or better than that of \textsc{StreamLS}. Meanwhile, the efficiency of SP-FSM is consistently higher than that of \textsc{StreamLS},
particularly so for larger values of $k$.
Thirdly, the performance of SP-FSM is hardly affected by the buffer size:
The solution quality and running time of SP-FSM are nearly identical when
setting the buffer size to be unlimited or $2k$. This confirms the effectiveness
of the buffer management strategies we propose.
Fourthly, the subsampling technique used in \textsc{StreamLS+S} does not perform well in our scenario:
although it obviously improves the efficiency upon \textsc{StreamLS},
its solution quality becomes significantly inferior to any other algorithm.

In the third set of experiments, we test the scalability of different algorithms with varying
the number $l$ of groups and the dataset size $n$ on SYN when $k$ is fixed to $500$.
Because the results for PR and ER are similar to each other, we only present the results for PR.
The performance of multi-pass streaming algorithms is shown in Figure~\ref{fig:mc-mp-nl}.
The solution utilities of different algorithms keep steady w.r.t.~$l$
while growing with increasing $n$ as expected.
Meanwhile, the solution quality of \textsc{Greedy}, \textsc{MP-Stream-LS}, and \textsc{MP-FSM}
is close to each other with varying $l$ and $n$. The difference in utilities are within
$ 5\% $ in all cases. Furthermore, the running time of all algorithms generally keeps steady
for different values of $l$ and grows near linearly with increasing $n$.
At the same time, \textsc{MP-FSM} runs nearly $10$ and $100$ times faster than \textsc{Greedy}
and \textsc{MP-Stream-LS}, respectively, for different values of $l$ and $n$.
The performance of single-pass streaming algorithms on SYN is shown in Figure~\ref{fig:mc-sp-nl}.
Since SP-FSM shows nearly identical performance for different buffer sizes,
we only present the results of SP-FSM ($k^\prime = 2k$) here.
Generally, we observe the same trends as the multi-pass case with varying $l$ and $n$.
For different values of $l$ and $n$,  the solution quality of SP-FSM and \textsc{StreamLS} is close to
each other, but SP-FSM runs much faster than \textsc{StreamLS}.
With the benefit of subsampling, \textsc{StreamLS+S} has much higher efficiency than
SP-FSM and \textsc{StreamLS}.
Nevertheless, its solution quality is obviously worse than them.

In summary, for maximum coverage on large graphs, our experimental results demonstrate that
our proposed algorithms MP-FSM and SP-FSM manage to pay small ``prices'' for the restrictions of the settings
(i.e., fairness constraint and streaming data access). And compared with the state-of-the-art algorithms,
they exhibit an excellent combination of performance in terms of running time and solution quality.

\subsection{Personalized Recommendation}\label{subsec:exp:pdm}

The personalized recommendation problem has been used for benchmarking submodular maximization algorithms
in~\cite{DBLP:conf/nips/MitrovicBNTC17,DBLP:conf/icml/Norouzi-FardTMZ18}.
Its goal is to select a subset $S$ of $k$ items that is both relevant to a given user $u$ and
well represents all items in the collection $V$.
Formally, each query user $u$ and each item $v$ in $V$ are denoted by feature vectors in $\mathbb{R}^d$.
The relevance between a user and an item is computed by the inner product of their feature vectors.
The objective function $f$ is defined as follows:
\begin{equation*}
  f(S) = \lambda \cdot \sum_{v^{\prime} \in V} \max_{v \in S} \langle v^{\prime},v \rangle
  + (1-\lambda) \cdot \sum_{v \in S} \langle u,v \rangle
\end{equation*}
and, again, $f$ is known to be nonnegative, monotone and submodular~\cite{DBLP:conf/nips/MitrovicBNTC17}.
The first term measures how well a subset $S$ represents the collection $V$;
the second term denotes the relevance of $S$ to user $u$;
and the parameter $\lambda$ trades off between both terms.
We set $\lambda=0.75$ following~\cite{DBLP:conf/nips/MitrovicBNTC17,DBLP:conf/icml/Norouzi-FardTMZ18}
in our experiments.

\begin{figure}[t]
  \includegraphics[width=0.475\textwidth]{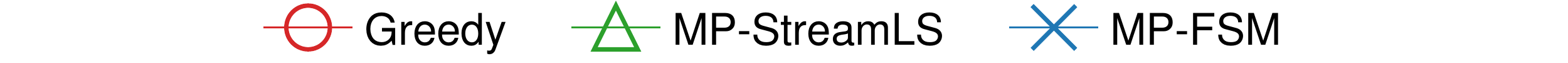}
  \\
  \includegraphics[width=0.23\textwidth]{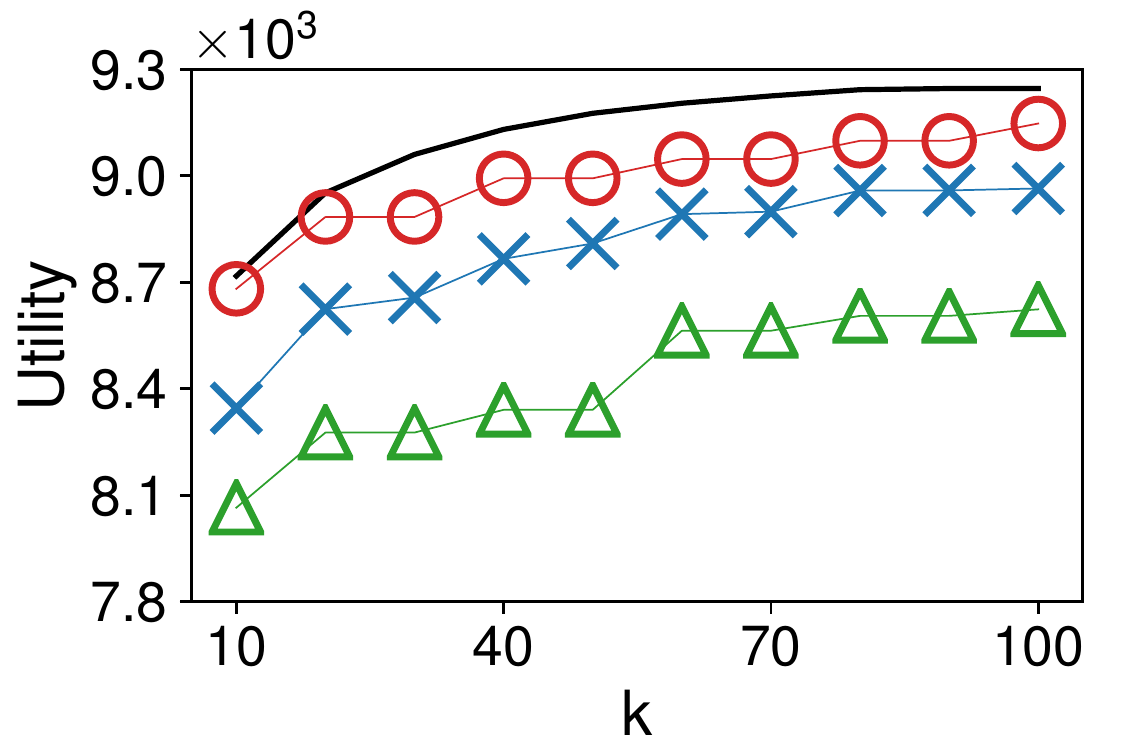}
  \includegraphics[width=0.23\textwidth]{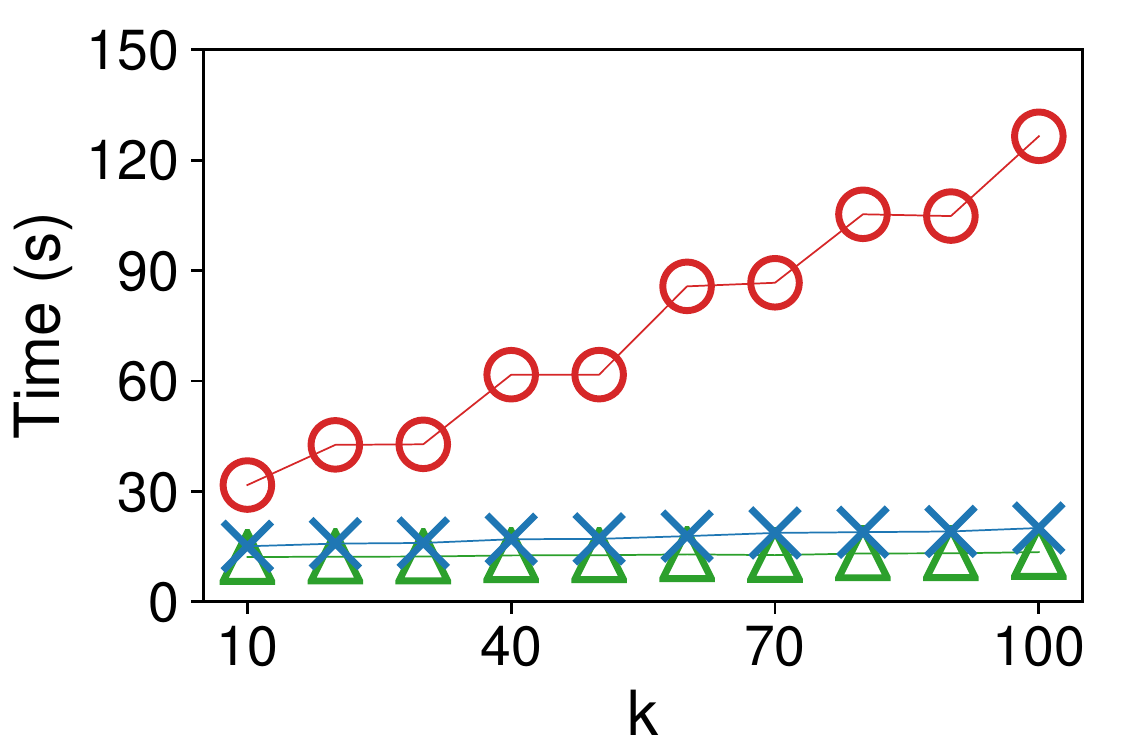}
  \caption{Performance of multi-pass algorithms on MovieLens. The utilities of \textsc{Greedy} without any fairness constraint are plotted as a black line.}
  \label{fig:pr-mp-k}
  \Description{recommendation, multi-pass, k}
\end{figure}
\begin{figure}[t]
  \includegraphics[width=0.475\textwidth]{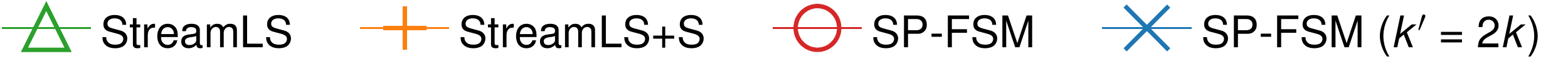}
  \\
  \includegraphics[width=0.23\textwidth]{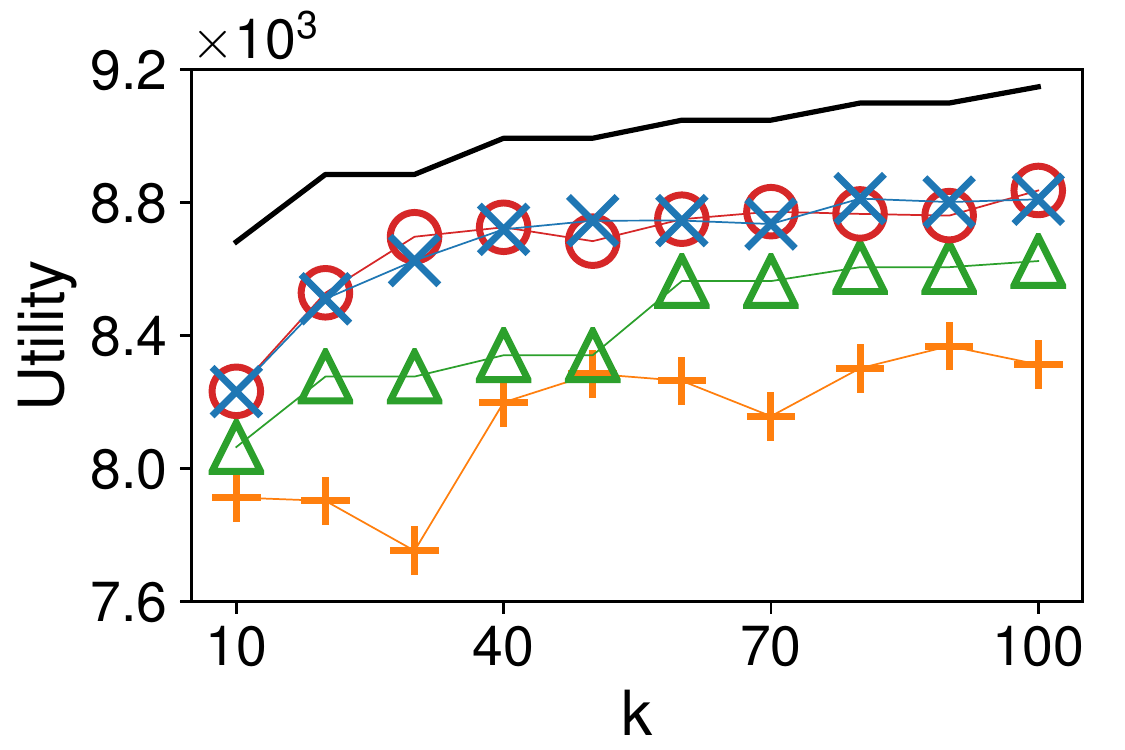}
  \includegraphics[width=0.23\textwidth]{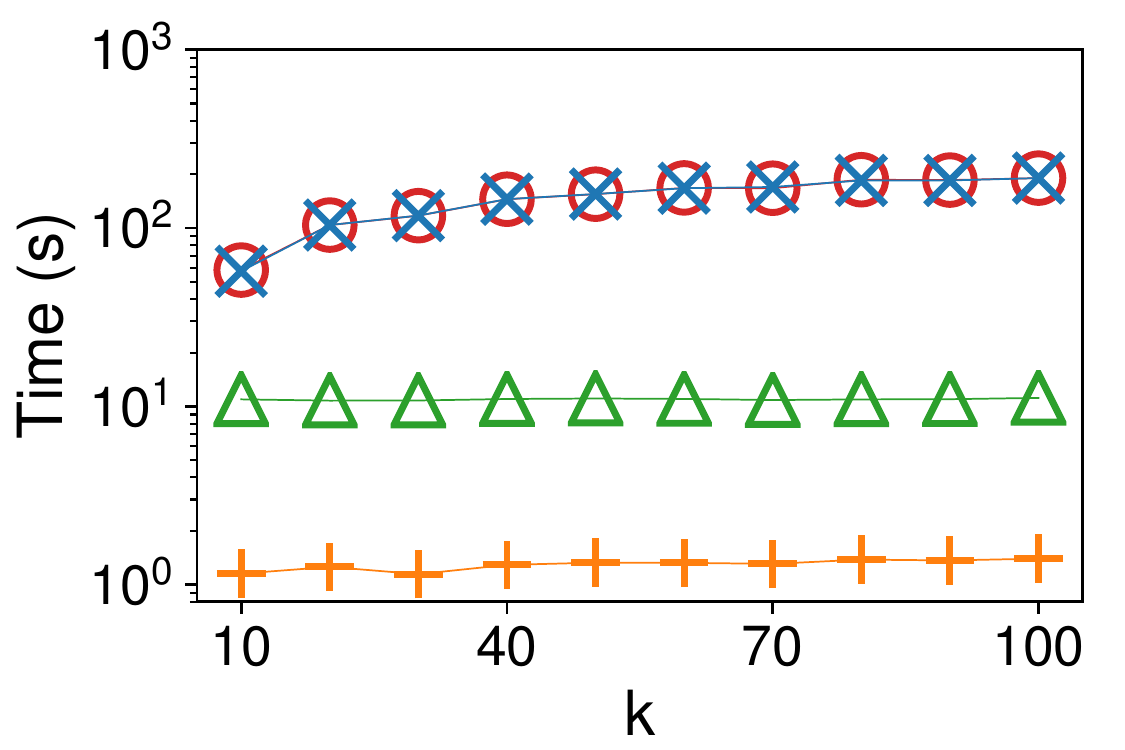}
  \caption{Performance of single-pass algorithms on MovieLens. The utilities of \textsc{Greedy} are plotted as a black line.}
  \label{fig:pr-sp-k}
  \Description{recommendation, single-pass, k}
\end{figure}

We perform the experiments for personalized recommendation on the \textbf{MovieLens}
dataset\footnote{\url{https://grouplens.org/datasets/movielens/}}.
It contains 3,883 items (movies) and 6,040 users with one million user ratings for movies.
We denote each item or user as a $50$-dimensional vector by performing
Nonnegative Matrix Factorization (NMF)~\cite{DBLP:journals/tkde/WangZ13} on the user-item rating matrix.
The items are partitioned into $l=10$ groups according to \emph{genre}.
Since the results for PR and ER are similar to each other,
we omit the results for ER in these experiments.

We present the performance of multi-pass streaming algorithms by ranging $k$ from $10$ to $100$
in Figure~\ref{fig:pr-mp-k}. Since the number $l$ of groups is relatively large compared to $k$,
the utility losses caused by fairness constraints are more significant than those in \emph{maximum coverage}.
Among all multi-pass streaming algorithms, \textsc{Greedy} runs the slowest
but achieves the best solution quality.
Moreover, MP-FSM shows higher efficiency than \textsc{Greedy}, especially when $k$ becomes larger.
Meanwhile, it provides solutions of at least $96\%$ utilities of the solutions of \textsc{Greedy}.
Although \textsc{MP-StreamLS} runs faster than MP-FSM and \textsc{Greedy}
because of fewer updates in solutions, its solution quality becomes worse as well.
We describe the performance of single-pass streaming algorithms by ranging $k$ from $10$ to $100$
in Figure~\ref{fig:pr-sp-k}. Similar to the case of \emph{maximum coverage}, the solution utilities
of SP-FSM (with unlimited and bounded buffer sizes) are around $10\%$ lower than
those of \textsc{Greedy} because only a single pass over the stream is permitted.
Nevertheless, SP-FSM provides solutions of higher quality than \textsc{StreamLS}
at the expense of longer running time.
Finally, \textsc{StreamLS+S} still brings great improvements in efficiency
but leads to obvious losses in solution quality.

In summary, for personalized recommendation, our experimental results demonstrate that
our proposed algorithms MP-FSM and SP-FSM have good performance compared with the state-of-the-art algorithms:
they provide solutions of higher quality than the local search based streaming algorithms
(i.e., \textsc{MP-StreamLS}, \textsc{StreamLS}, and \textsc{StreamLS+S}) at the expense of lower efficiency.

%!TEX root = main.tex
\section{Conclusion}\label{sec:conclusion}

In this paper, we studied the problem of extracting fair and representative items from data streams.
We formulated the problem as maximizing monotone submodular functions subject to partition matroid constraints.
We first proposed a $ (\frac{1}{2}-\varepsilon) $-approximation multi-pass streaming algorithm called MP-FSM
for the problem. Then, we designed a single-pass streaming algorithm called SP-FSM for the problem.
SP-FSM had the same approximation ratio of $ (\frac{1}{2}-\varepsilon) $ as MP-FSM when an unlimited buffer size is permitted, which improved the best-known approximation ratio of $ \frac{1}{4} $ in the literature.
We further considered the practical implementation of SP-FSM when the buffer sizes are bounded.
Finally, extensive experimental results on two real-world applications confirmed the efficiency, effectiveness, and scalability of our proposed algorithms.

\begin{acks}
  We thank the anonymous reviewers for their helpful comments to improve this paper.
  Yanhao Wang and Michael Mathioudakis have been supported by the MLDB project of Academy of Finland
  (decision number: 322046). 
  Francesco Fabbri has been supported by the Helsinki Institute for Information Technology (HIIT).
\end{acks}

\bibliographystyle{ACM-Reference-Format}
\bibliography{references}


\begin{thebibliography}{40}

%%% the Web address.

\ifx \showCODEN    \undefined \def \showCODEN     #1{\unskip}     \fi
\ifx \showDOI      \undefined \def \showDOI       #1{#1}\fi
\ifx \showISBNx    \undefined \def \showISBNx     #1{\unskip}     \fi
\ifx \showISBNxiii \undefined \def \showISBNxiii  #1{\unskip}     \fi
\ifx \showISSN     \undefined \def \showISSN      #1{\unskip}     \fi
\ifx \showLCCN     \undefined \def \showLCCN      #1{\unskip}     \fi
\ifx \shownote     \undefined \def \shownote      #1{#1}          \fi
\ifx \showarticletitle \undefined \def \showarticletitle #1{#1}   \fi
\ifx \showURL      \undefined \def \showURL       {\relax}        \fi
% invisible to TeX
\providecommand\bibfield[2]{#2}
\providecommand\bibinfo[2]{#2}
\providecommand\natexlab[1]{#1}
\providecommand\showeprint[2][]{arXiv:#2}

\bibitem[\protect\citeauthoryear{Abbassi, Mirrokni, and Thakur}{Abbassi
  et~al\mbox{.}}{2013}]%
        {DBLP:conf/kdd/AbbassiMT13}
\bibfield{author}{\bibinfo{person}{Zeinab Abbassi}, \bibinfo{person}{Vahab~S.
  Mirrokni}, {and} \bibinfo{person}{Mayur Thakur}.}
  \bibinfo{year}{2013}\natexlab{}.
\newblock \showarticletitle{Diversity maximization under matroid constraints}.
  In \bibinfo{booktitle}{\emph{{KDD}}}. \bibinfo{pages}{32--40}.
\newblock


\bibitem[\protect\citeauthoryear{Alaluf, Ene, Feldman, Nguyen, and Suh}{Alaluf
  et~al\mbox{.}}{2020}]%
        {DBLP:conf/icalp/AlalufEFNS20}
\bibfield{author}{\bibinfo{person}{Naor Alaluf}, \bibinfo{person}{Alina Ene},
  \bibinfo{person}{Moran Feldman}, \bibinfo{person}{Huy~L. Nguyen}, {and}
  \bibinfo{person}{Andrew Suh}.} \bibinfo{year}{2020}\natexlab{}.
\newblock \showarticletitle{Optimal Streaming Algorithms for Submodular
  Maximization with Cardinality Constraints}. In
  \bibinfo{booktitle}{\emph{{ICALP}}}. \bibinfo{pages}{6:1--6:19}.
\newblock


\bibitem[\protect\citeauthoryear{Albert and Barab\'asi}{Albert and
  Barab\'asi}{2002}]%
        {RevModPhys.74.47}
\bibfield{author}{\bibinfo{person}{R\'eka Albert} {and}
  \bibinfo{person}{Albert-L\'aszl\'o Barab\'asi}.}
  \bibinfo{year}{2002}\natexlab{}.
\newblock \showarticletitle{Statistical mechanics of complex networks}.
\newblock \bibinfo{journal}{\emph{Rev. Mod. Phys.}}  \bibinfo{volume}{74}
  (\bibinfo{year}{2002}), \bibinfo{pages}{47--97}.
\newblock
Issue 1.


\bibitem[\protect\citeauthoryear{Badanidiyuru, Mirzasoleiman, Karbasi, and
  Krause}{Badanidiyuru et~al\mbox{.}}{2014}]%
        {DBLP:conf/kdd/BadanidiyuruMKK14}
\bibfield{author}{\bibinfo{person}{Ashwinkumar Badanidiyuru},
  \bibinfo{person}{Baharan Mirzasoleiman}, \bibinfo{person}{Amin Karbasi},
  {and} \bibinfo{person}{Andreas Krause}.} \bibinfo{year}{2014}\natexlab{}.
\newblock \showarticletitle{Streaming Submodular Maximization: Massive Data
  Summarization on the Fly}. In \bibinfo{booktitle}{\emph{{KDD}}}.
  \bibinfo{pages}{671--680}.
\newblock


\bibitem[\protect\citeauthoryear{Celis, Keswani, Straszak, Deshpande, Kathuria,
  and Vishnoi}{Celis et~al\mbox{.}}{2018}]%
        {DBLP:conf/icml/CelisKS0KV18}
\bibfield{author}{\bibinfo{person}{L.~Elisa Celis}, \bibinfo{person}{Vijay
  Keswani}, \bibinfo{person}{Damian Straszak}, \bibinfo{person}{Amit
  Deshpande}, \bibinfo{person}{Tarun Kathuria}, {and}
  \bibinfo{person}{Nisheeth~K. Vishnoi}.} \bibinfo{year}{2018}\natexlab{}.
\newblock \showarticletitle{Fair and Diverse DPP-Based Data Summarization}. In
  \bibinfo{booktitle}{\emph{{ICML}}}. \bibinfo{pages}{715--724}.
\newblock


\bibitem[\protect\citeauthoryear{Chakrabarti and Kale}{Chakrabarti and
  Kale}{2015}]%
        {DBLP:journals/mp/ChakrabartiK15}
\bibfield{author}{\bibinfo{person}{Amit Chakrabarti} {and}
  \bibinfo{person}{Sagar Kale}.} \bibinfo{year}{2015}\natexlab{}.
\newblock \showarticletitle{Submodular maximization meets streaming: matchings,
  matroids, and more}.
\newblock \bibinfo{journal}{\emph{Math. Program.}} \bibinfo{volume}{154},
  \bibinfo{number}{1-2} (\bibinfo{year}{2015}), \bibinfo{pages}{225--247}.
\newblock


\bibitem[\protect\citeauthoryear{Chan, Huang, Jiang, Kang, and Tang}{Chan
  et~al\mbox{.}}{2017}]%
        {DBLP:conf/soda/Chan0JKT17}
\bibfield{author}{\bibinfo{person}{T.{-}H.~Hubert Chan}, \bibinfo{person}{Zhiyi
  Huang}, \bibinfo{person}{Shaofeng~H.{-}C. Jiang}, \bibinfo{person}{Ning
  Kang}, {and} \bibinfo{person}{Zhihao~Gavin Tang}.}
  \bibinfo{year}{2017}\natexlab{}.
\newblock \showarticletitle{Online Submodular Maximization with Free Disposal:
  Randomization Beats {\textonequarter} for Partition Matroids}. In
  \bibinfo{booktitle}{\emph{{SODA}}}. \bibinfo{pages}{1204--1223}.
\newblock


\bibitem[\protect\citeauthoryear{Chekuri, Gupta, and Quanrud}{Chekuri
  et~al\mbox{.}}{2015}]%
        {DBLP:conf/icalp/ChekuriGQ15}
\bibfield{author}{\bibinfo{person}{Chandra Chekuri}, \bibinfo{person}{Shalmoli
  Gupta}, {and} \bibinfo{person}{Kent Quanrud}.}
  \bibinfo{year}{2015}\natexlab{}.
\newblock \showarticletitle{Streaming Algorithms for Submodular Function
  Maximization}. In \bibinfo{booktitle}{\emph{{ICALP}}}.
  \bibinfo{pages}{318--330}.
\newblock


\bibitem[\protect\citeauthoryear{Chiplunkar, Kale, and Ramamoorthy}{Chiplunkar
  et~al\mbox{.}}{2020}]%
        {chiplunkar2020solve}
\bibfield{author}{\bibinfo{person}{Ashish Chiplunkar}, \bibinfo{person}{Sagar
  Kale}, {and} \bibinfo{person}{Sivaramakrishnan~Natarajan Ramamoorthy}.}
  \bibinfo{year}{2020}\natexlab{}.
\newblock \showarticletitle{How to Solve Fair $k$-Center in Massive Data
  Models}. In \bibinfo{booktitle}{\emph{{ICML}}}. \bibinfo{pages}{6887--6896}.
\newblock


\bibitem[\protect\citeauthoryear{Chouldechova and Roth}{Chouldechova and
  Roth}{2020}]%
        {DBLP:journals/cacm/ChouldechovaR20}
\bibfield{author}{\bibinfo{person}{Alexandra Chouldechova} {and}
  \bibinfo{person}{Aaron Roth}.} \bibinfo{year}{2020}\natexlab{}.
\newblock \showarticletitle{A snapshot of the frontiers of fairness in machine
  learning}.
\newblock \bibinfo{journal}{\emph{Commun. {ACM}}} \bibinfo{volume}{63},
  \bibinfo{number}{5} (\bibinfo{year}{2020}), \bibinfo{pages}{82--89}.
\newblock


\bibitem[\protect\citeauthoryear{Dash, Shandilya, Biswas, Ghosh, Ghosh, and
  Chakraborty}{Dash et~al\mbox{.}}{2019}]%
        {DBLP:journals/pacmhci/DashSBGGC19}
\bibfield{author}{\bibinfo{person}{Abhisek Dash}, \bibinfo{person}{Anurag
  Shandilya}, \bibinfo{person}{Arindam Biswas}, \bibinfo{person}{Kripabandhu
  Ghosh}, \bibinfo{person}{Saptarshi Ghosh}, {and} \bibinfo{person}{Abhijnan
  Chakraborty}.} \bibinfo{year}{2019}\natexlab{}.
\newblock \showarticletitle{Summarizing User-generated Textual Content:
  Motivation and Methods for Fairness in Algorithmic Summaries}.
\newblock \bibinfo{journal}{\emph{Proc. ACM Hum. Comput. Interact.}}
  \bibinfo{volume}{3}, \bibinfo{number}{CSCW} (\bibinfo{year}{2019}),
  \bibinfo{pages}{172:1--172:28}.
\newblock


\bibitem[\protect\citeauthoryear{Epasto, Lattanzi, Vassilvitskii, and
  Zadimoghaddam}{Epasto et~al\mbox{.}}{2017}]%
        {DBLP:conf/www/EpastoLVZ17}
\bibfield{author}{\bibinfo{person}{Alessandro Epasto}, \bibinfo{person}{Silvio
  Lattanzi}, \bibinfo{person}{Sergei Vassilvitskii}, {and}
  \bibinfo{person}{Morteza Zadimoghaddam}.} \bibinfo{year}{2017}\natexlab{}.
\newblock \showarticletitle{Submodular Optimization Over Sliding Windows}. In
  \bibinfo{booktitle}{\emph{{WWW}}}. \bibinfo{pages}{421--430}.
\newblock


\bibitem[\protect\citeauthoryear{Feige}{Feige}{1998}]%
        {DBLP:journals/jacm/Feige98}
\bibfield{author}{\bibinfo{person}{Uriel Feige}.}
  \bibinfo{year}{1998}\natexlab{}.
\newblock \showarticletitle{A Threshold of ln \emph{n} for Approximating Set
  Cover}.
\newblock \bibinfo{journal}{\emph{J. {ACM}}} \bibinfo{volume}{45},
  \bibinfo{number}{4} (\bibinfo{year}{1998}), \bibinfo{pages}{634--652}.
\newblock


\bibitem[\protect\citeauthoryear{Feldman, Karbasi, and Kazemi}{Feldman
  et~al\mbox{.}}{2018}]%
        {DBLP:conf/nips/FeldmanK018}
\bibfield{author}{\bibinfo{person}{Moran Feldman}, \bibinfo{person}{Amin
  Karbasi}, {and} \bibinfo{person}{Ehsan Kazemi}.}
  \bibinfo{year}{2018}\natexlab{}.
\newblock \showarticletitle{Do Less, Get More: Streaming Submodular
  Maximization with Subsampling}. In \bibinfo{booktitle}{\emph{{NeurIPS}}}.
  \bibinfo{pages}{730--740}.
\newblock


\bibitem[\protect\citeauthoryear{Fisher, Nemhauser, and Wolsey}{Fisher
  et~al\mbox{.}}{1978}]%
        {Fisher1978}
\bibfield{author}{\bibinfo{person}{Marshall~L. Fisher},
  \bibinfo{person}{George~L. Nemhauser}, {and} \bibinfo{person}{Laurence~A.
  Wolsey}.} \bibinfo{year}{1978}\natexlab{}.
\newblock \showarticletitle{An analysis of approximations for maximizing
  submodular set functions---{II}}.
\newblock In \bibinfo{booktitle}{\emph{Polyhedral Combinatorics}},
  \bibfield{editor}{\bibinfo{person}{Michel~L. Balinski} {and}
  \bibinfo{person}{Alan~J. Hoffman}} (Eds.). \bibinfo{publisher}{Springer
  Berlin Heidelberg}, \bibinfo{pages}{73--87}.
\newblock
\showISBNx{978-3-642-00790-3}


\bibitem[\protect\citeauthoryear{Galbrun, Gionis, and Tatti}{Galbrun
  et~al\mbox{.}}{2014}]%
        {DBLP:journals/datamine/GalbrunGT14}
\bibfield{author}{\bibinfo{person}{Esther Galbrun}, \bibinfo{person}{Aristides
  Gionis}, {and} \bibinfo{person}{Nikolaj Tatti}.}
  \bibinfo{year}{2014}\natexlab{}.
\newblock \showarticletitle{Overlapping community detection in labeled graphs}.
\newblock \bibinfo{journal}{\emph{Data Min. Knowl. Discov.}}
  \bibinfo{volume}{28}, \bibinfo{number}{5-6} (\bibinfo{year}{2014}),
  \bibinfo{pages}{1586--1610}.
\newblock


\bibitem[\protect\citeauthoryear{Gomes and Krause}{Gomes and Krause}{2010}]%
        {DBLP:conf/icml/GomesK10}
\bibfield{author}{\bibinfo{person}{Ryan Gomes} {and} \bibinfo{person}{Andreas
  Krause}.} \bibinfo{year}{2010}\natexlab{}.
\newblock \showarticletitle{Budgeted Nonparametric Learning from Data Streams}.
  In \bibinfo{booktitle}{\emph{{ICML}}}. \bibinfo{pages}{391–398}.
\newblock


\bibitem[\protect\citeauthoryear{Huang, Thiery, and Ward}{Huang
  et~al\mbox{.}}{2020}]%
        {DBLP:conf/approx/HuangTW20}
\bibfield{author}{\bibinfo{person}{Chien{-}Chung Huang},
  \bibinfo{person}{Theophile Thiery}, {and} \bibinfo{person}{Justin Ward}.}
  \bibinfo{year}{2020}\natexlab{}.
\newblock \showarticletitle{Improved Multi-Pass Streaming Algorithms for
  Submodular Maximization with Matroid Constraints}. In
  \bibinfo{booktitle}{\emph{{APPROX/RANDOM}}}. \bibinfo{pages}{62:1--62:19}.
\newblock


\bibitem[\protect\citeauthoryear{Jones, L{\^e}~Nguy{\^e}n, and Nguyen}{Jones
  et~al\mbox{.}}{2020}]%
        {DBLP:conf/icml/JonesNN20}
\bibfield{author}{\bibinfo{person}{Matthew Jones}, \bibinfo{person}{Huy
  L{\^e}~Nguy{\^e}n}, {and} \bibinfo{person}{Thy Nguyen}.}
  \bibinfo{year}{2020}\natexlab{}.
\newblock \showarticletitle{Fair k-Centers via Maximum Matching}. In
  \bibinfo{booktitle}{\emph{{ICML}}}. \bibinfo{pages}{7460--7469}.
\newblock


\bibitem[\protect\citeauthoryear{Kay, Matuszek, and Munson}{Kay
  et~al\mbox{.}}{2015}]%
        {DBLP:conf/chi/KayMM15}
\bibfield{author}{\bibinfo{person}{Matthew Kay}, \bibinfo{person}{Cynthia
  Matuszek}, {and} \bibinfo{person}{Sean~A. Munson}.}
  \bibinfo{year}{2015}\natexlab{}.
\newblock \showarticletitle{Unequal Representation and Gender Stereotypes in
  Image Search Results for Occupations}. In \bibinfo{booktitle}{\emph{{CHI}}}.
  \bibinfo{pages}{3819--3828}.
\newblock


\bibitem[\protect\citeauthoryear{Kazemi, Mitrovic, Zadimoghaddam, Lattanzi, and
  Karbasi}{Kazemi et~al\mbox{.}}{2019}]%
        {DBLP:conf/icml/0001MZLK19}
\bibfield{author}{\bibinfo{person}{Ehsan Kazemi}, \bibinfo{person}{Marko
  Mitrovic}, \bibinfo{person}{Morteza Zadimoghaddam}, \bibinfo{person}{Silvio
  Lattanzi}, {and} \bibinfo{person}{Amin Karbasi}.}
  \bibinfo{year}{2019}\natexlab{}.
\newblock \showarticletitle{Submodular Streaming in All Its Glory: Tight
  Approximation, Minimum Memory and Low Adaptive Complexity}. In
  \bibinfo{booktitle}{\emph{{ICML}}}. \bibinfo{pages}{3311--3320}.
\newblock


\bibitem[\protect\citeauthoryear{Kazemi, Zadimoghaddam, and Karbasi}{Kazemi
  et~al\mbox{.}}{2018}]%
        {DBLP:conf/icml/0001ZK18}
\bibfield{author}{\bibinfo{person}{Ehsan Kazemi}, \bibinfo{person}{Morteza
  Zadimoghaddam}, {and} \bibinfo{person}{Amin Karbasi}.}
  \bibinfo{year}{2018}\natexlab{}.
\newblock \showarticletitle{Scalable Deletion-Robust Submodular Maximization:
  Data Summarization with Privacy and Fairness Constraints}. In
  \bibinfo{booktitle}{\emph{{ICML}}}. \bibinfo{pages}{2549--2558}.
\newblock


\bibitem[\protect\citeauthoryear{Kempe, Kleinberg, and Tardos}{Kempe
  et~al\mbox{.}}{2003}]%
        {DBLP:conf/kdd/KempeKT03}
\bibfield{author}{\bibinfo{person}{David Kempe}, \bibinfo{person}{Jon~M.
  Kleinberg}, {and} \bibinfo{person}{{\'{E}}va Tardos}.}
  \bibinfo{year}{2003}\natexlab{}.
\newblock \showarticletitle{Maximizing the spread of influence through a social
  network}. In \bibinfo{booktitle}{\emph{{KDD}}}. \bibinfo{pages}{137--146}.
\newblock


\bibitem[\protect\citeauthoryear{Kleindessner, Awasthi, and
  Morgenstern}{Kleindessner et~al\mbox{.}}{2019}]%
        {DBLP:conf/icml/KleindessnerAM19}
\bibfield{author}{\bibinfo{person}{Matth{\"{a}}us Kleindessner},
  \bibinfo{person}{Pranjal Awasthi}, {and} \bibinfo{person}{Jamie
  Morgenstern}.} \bibinfo{year}{2019}\natexlab{}.
\newblock \showarticletitle{Fair k-Center Clustering for Data Summarization}.
  In \bibinfo{booktitle}{\emph{{ICML}}}. \bibinfo{pages}{3448--3457}.
\newblock


\bibitem[\protect\citeauthoryear{Krause and Golovin}{Krause and
  Golovin}{2014}]%
        {DBLP:books/cu/p/0001G14}
\bibfield{author}{\bibinfo{person}{Andreas Krause} {and}
  \bibinfo{person}{Daniel Golovin}.} \bibinfo{year}{2014}\natexlab{}.
\newblock \showarticletitle{Submodular Function Maximization}.
\newblock In \bibinfo{booktitle}{\emph{Tractability: Practical Approaches to
  Hard Problems}}, \bibfield{editor}{\bibinfo{person}{Lucas Bordeaux},
  \bibinfo{person}{Youssef Hamadi}, {and} \bibinfo{person}{Pushmeet Kohli}}
  (Eds.). \bibinfo{publisher}{Cambridge University Press},
  \bibinfo{pages}{71--104}.
\newblock


\bibitem[\protect\citeauthoryear{Kumar, Moseley, Vassilvitskii, and
  Vattani}{Kumar et~al\mbox{.}}{2015}]%
        {DBLP:journals/topc/KumarMVV15}
\bibfield{author}{\bibinfo{person}{Ravi Kumar}, \bibinfo{person}{Benjamin
  Moseley}, \bibinfo{person}{Sergei Vassilvitskii}, {and}
  \bibinfo{person}{Andrea Vattani}.} \bibinfo{year}{2015}\natexlab{}.
\newblock \showarticletitle{Fast Greedy Algorithms in {MapReduce} and
  Streaming}.
\newblock \bibinfo{journal}{\emph{{ACM} Trans. Parallel Comput.}}
  \bibinfo{volume}{2}, \bibinfo{number}{3} (\bibinfo{year}{2015}),
  \bibinfo{pages}{14:1--14:22}.
\newblock


\bibitem[\protect\citeauthoryear{Leskovec, Krause, Guestrin, Faloutsos, {Van
  Briesen}, and Glance}{Leskovec et~al\mbox{.}}{2007}]%
        {DBLP:conf/kdd/LeskovecKGFVG07}
\bibfield{author}{\bibinfo{person}{Jure Leskovec}, \bibinfo{person}{Andreas
  Krause}, \bibinfo{person}{Carlos Guestrin}, \bibinfo{person}{Christos
  Faloutsos}, \bibinfo{person}{Jeanne~M. {Van Briesen}}, {and}
  \bibinfo{person}{Natalie~S. Glance}.} \bibinfo{year}{2007}\natexlab{}.
\newblock \showarticletitle{Cost-effective outbreak detection in networks}. In
  \bibinfo{booktitle}{\emph{{KDD}}}. \bibinfo{pages}{420--429}.
\newblock


\bibitem[\protect\citeauthoryear{Lindgren, Wu, and Dimakis}{Lindgren
  et~al\mbox{.}}{2016}]%
        {DBLP:conf/nips/LindgrenWD16}
\bibfield{author}{\bibinfo{person}{Erik~M. Lindgren}, \bibinfo{person}{Shanshan
  Wu}, {and} \bibinfo{person}{Alexandros~G. Dimakis}.}
  \bibinfo{year}{2016}\natexlab{}.
\newblock \showarticletitle{Leveraging Sparsity for Efficient Submodular Data
  Summarization}. In \bibinfo{booktitle}{\emph{{NIPS}}}.
  \bibinfo{pages}{3414--3422}.
\newblock


\bibitem[\protect\citeauthoryear{Mirzasoleiman, Karbasi, and
  Krause}{Mirzasoleiman et~al\mbox{.}}{2017}]%
        {DBLP:conf/icml/MirzasoleimanK017}
\bibfield{author}{\bibinfo{person}{Baharan Mirzasoleiman},
  \bibinfo{person}{Amin Karbasi}, {and} \bibinfo{person}{Andreas Krause}.}
  \bibinfo{year}{2017}\natexlab{}.
\newblock \showarticletitle{Deletion-Robust Submodular Maximization: Data
  Summarization with "the Right to be Forgotten"}. In
  \bibinfo{booktitle}{\emph{{ICML}}}. \bibinfo{pages}{2449--2458}.
\newblock


\bibitem[\protect\citeauthoryear{Mitrovic, Bogunovic, Norouzi{-}Fard,
  Tarnawski, and Cevher}{Mitrovic et~al\mbox{.}}{2017}]%
        {DBLP:conf/nips/MitrovicBNTC17}
\bibfield{author}{\bibinfo{person}{Slobodan Mitrovic}, \bibinfo{person}{Ilija
  Bogunovic}, \bibinfo{person}{Ashkan Norouzi{-}Fard}, \bibinfo{person}{Jakub
  Tarnawski}, {and} \bibinfo{person}{Volkan Cevher}.}
  \bibinfo{year}{2017}\natexlab{}.
\newblock \showarticletitle{Streaming Robust Submodular Maximization: {A}
  Partitioned Thresholding Approach}. In \bibinfo{booktitle}{\emph{{NIPS}}}.
  \bibinfo{pages}{4557--4566}.
\newblock


\bibitem[\protect\citeauthoryear{Nemhauser, Wolsey, and Fisher}{Nemhauser
  et~al\mbox{.}}{1978}]%
        {DBLP:journals/mp/NemhauserWF78}
\bibfield{author}{\bibinfo{person}{George~L. Nemhauser},
  \bibinfo{person}{Laurence~A. Wolsey}, {and} \bibinfo{person}{Marshall~L.
  Fisher}.} \bibinfo{year}{1978}\natexlab{}.
\newblock \showarticletitle{An analysis of approximations for maximizing
  submodular set functions---{I}}.
\newblock \bibinfo{journal}{\emph{Math. Program.}} \bibinfo{volume}{14},
  \bibinfo{number}{1} (\bibinfo{year}{1978}), \bibinfo{pages}{265--294}.
\newblock


\bibitem[\protect\citeauthoryear{Norouzi{-}Fard, Tarnawski, Mitrovic, Zandieh,
  Mousavifar, and Svensson}{Norouzi{-}Fard et~al\mbox{.}}{2018}]%
        {DBLP:conf/icml/Norouzi-FardTMZ18}
\bibfield{author}{\bibinfo{person}{Ashkan Norouzi{-}Fard},
  \bibinfo{person}{Jakub Tarnawski}, \bibinfo{person}{Slobodan Mitrovic},
  \bibinfo{person}{Amir Zandieh}, \bibinfo{person}{Aidasadat Mousavifar}, {and}
  \bibinfo{person}{Ola Svensson}.} \bibinfo{year}{2018}\natexlab{}.
\newblock \showarticletitle{Beyond 1/2-Approximation for Submodular
  Maximization on Massive Data Streams}. In \bibinfo{booktitle}{\emph{{ICML}}}.
  \bibinfo{pages}{3826--3835}.
\newblock


\bibitem[\protect\citeauthoryear{Saha and Getoor}{Saha and Getoor}{2009}]%
        {DBLP:conf/sdm/SahaG09}
\bibfield{author}{\bibinfo{person}{Barna Saha} {and} \bibinfo{person}{Lise
  Getoor}.} \bibinfo{year}{2009}\natexlab{}.
\newblock \showarticletitle{On Maximum Coverage in the Streaming Model {\&}
  Application to Multi-topic Blog-Watch}. In \bibinfo{booktitle}{\emph{{SDM}}}.
  \bibinfo{pages}{697--708}.
\newblock


\bibitem[\protect\citeauthoryear{Serbos, Qi, Mamoulis, Pitoura, and
  Tsaparas}{Serbos et~al\mbox{.}}{2017}]%
        {DBLP:conf/www/SerbosQMPT17}
\bibfield{author}{\bibinfo{person}{Dimitris Serbos}, \bibinfo{person}{Shuyao
  Qi}, \bibinfo{person}{Nikos Mamoulis}, \bibinfo{person}{Evaggelia Pitoura},
  {and} \bibinfo{person}{Panayiotis Tsaparas}.}
  \bibinfo{year}{2017}\natexlab{}.
\newblock \showarticletitle{Fairness in Package-to-Group Recommendations}. In
  \bibinfo{booktitle}{\emph{{WWW}}}. \bibinfo{pages}{371--379}.
\newblock


\bibitem[\protect\citeauthoryear{Stoica, Han, and Chaintreau}{Stoica
  et~al\mbox{.}}{2020}]%
        {DBLP:conf/www/StoicaHC20}
\bibfield{author}{\bibinfo{person}{Ana{-}Andreea Stoica},
  \bibinfo{person}{Jessy~Xinyi Han}, {and} \bibinfo{person}{Augustin
  Chaintreau}.} \bibinfo{year}{2020}\natexlab{}.
\newblock \showarticletitle{Seeding Network Influence in Biased Networks and
  the Benefits of Diversity}. In \bibinfo{booktitle}{\emph{{WWW}}}.
  \bibinfo{pages}{2089--2098}.
\newblock


\bibitem[\protect\citeauthoryear{Vitter}{Vitter}{1985}]%
        {DBLP:journals/toms/Vitter85}
\bibfield{author}{\bibinfo{person}{Jeffrey~Scott Vitter}.}
  \bibinfo{year}{1985}\natexlab{}.
\newblock \showarticletitle{Random Sampling with a Reservoir}.
\newblock \bibinfo{journal}{\emph{{ACM} Trans. Math. Softw.}}
  \bibinfo{volume}{11}, \bibinfo{number}{1} (\bibinfo{year}{1985}),
  \bibinfo{pages}{37--57}.
\newblock


\bibitem[\protect\citeauthoryear{Wang, Fan, Li, and Tan}{Wang
  et~al\mbox{.}}{2017}]%
        {DBLP:journals/pvldb/WangFLT17}
\bibfield{author}{\bibinfo{person}{Yanhao Wang}, \bibinfo{person}{Qi Fan},
  \bibinfo{person}{Yuchen Li}, {and} \bibinfo{person}{Kian{-}Lee Tan}.}
  \bibinfo{year}{2017}\natexlab{}.
\newblock \showarticletitle{Real-Time Influence Maximization on Dynamic Social
  Streams}.
\newblock \bibinfo{journal}{\emph{Proc. {VLDB} Endow.}} \bibinfo{volume}{10},
  \bibinfo{number}{7} (\bibinfo{year}{2017}), \bibinfo{pages}{805--816}.
\newblock


\bibitem[\protect\citeauthoryear{Wang, Li, and Tan}{Wang et~al\mbox{.}}{2019}]%
        {DBLP:journals/tkde/WangLT19}
\bibfield{author}{\bibinfo{person}{Yanhao Wang}, \bibinfo{person}{Yuchen Li},
  {and} \bibinfo{person}{Kian{-}Lee Tan}.} \bibinfo{year}{2019}\natexlab{}.
\newblock \showarticletitle{Efficient Representative Subset Selection over
  Sliding Windows}.
\newblock \bibinfo{journal}{\emph{IEEE Trans. Knowl. Data Eng.}}
  \bibinfo{volume}{31}, \bibinfo{number}{7} (\bibinfo{year}{2019}),
  \bibinfo{pages}{1327--1340}.
\newblock


\bibitem[\protect\citeauthoryear{Wang and Zhang}{Wang and Zhang}{2013}]%
        {DBLP:journals/tkde/WangZ13}
\bibfield{author}{\bibinfo{person}{Yu{-}Xiong Wang} {and}
  \bibinfo{person}{Yu{-}Jin Zhang}.} \bibinfo{year}{2013}\natexlab{}.
\newblock \showarticletitle{Nonnegative Matrix Factorization: {A} Comprehensive
  Review}.
\newblock \bibinfo{journal}{\emph{{IEEE} Trans. Knowl. Data Eng.}}
  \bibinfo{volume}{25}, \bibinfo{number}{6} (\bibinfo{year}{2013}),
  \bibinfo{pages}{1336--1353}.
\newblock


\bibitem[\protect\citeauthoryear{Zhao, Shang, Wang, Lui, and Zhang}{Zhao
  et~al\mbox{.}}{2019}]%
        {DBLP:conf/aaai/ZhaoSWLZ19}
\bibfield{author}{\bibinfo{person}{Junzhou Zhao}, \bibinfo{person}{Shuo Shang},
  \bibinfo{person}{Pinghui Wang}, \bibinfo{person}{John C.~S. Lui}, {and}
  \bibinfo{person}{Xiangliang Zhang}.} \bibinfo{year}{2019}\natexlab{}.
\newblock \showarticletitle{Submodular Optimization over Streams with
  Inhomogeneous Decays}. In \bibinfo{booktitle}{\emph{{AAAI}}}.
  \bibinfo{pages}{5861--5868}.
\newblock


\end{thebibliography}

\end{document}